\documentclass{article}
\usepackage[page]{appendix}
\usepackage{xpatch}
\xpretocmd{\appendixpagename}{\sffamily}{}{}
\usepackage[utf8]{inputenc} 
\usepackage[T1]{fontenc}    
\usepackage{hyperref}      
\usepackage{booktabs}     
\usepackage{amsfonts}       
\usepackage{nicefrac}       
\usepackage{microtype}      
\usepackage{lipsum}		
\usepackage{graphicx}
\usepackage{doi}
\usepackage{amsmath}
\usepackage{amssymb}
\usepackage{multirow}
\usepackage[a4paper]{geometry}
\usepackage{float}
\usepackage{subcaption}
\usepackage{booktabs}
\usepackage{verbatim}
\usepackage[inline]{enumitem}
\usepackage{adjustbox}
\usepackage{graphicx}
\usepackage[dvipsnames]{xcolor}
\usepackage{natbib}
\usepackage{cite}
\usepackage{xr-hyper}
\usepackage{hyperref}
\providecommand{\keywords}[1]{\textbf{\textit{Keywords:}} #1}

\title{Fast geostatistical inference under positional uncertainty: Analysing DHS household survey data}

\date{}
\author{Umut Altay\qquad John Paige\qquad Andrea Riebler\qquad \\Geir-Arne Fuglstad\\ \\Department of Mathematical Sciences, Norwegian University \\of Science and Technology, Trondheim, Norway}

\begin{document}
\maketitle

\begin{abstract}
Household survey data from the Demographic and Health Surveys (DHS) Program is published with GPS coordinates. However, almost all geostatistical analyses of such data ignore that the published GPS coordinates are randomly displaced (jittered). In this short report, we develop a geostatistical model that accounts for the positional uncertainty when analysing DHS surveys, and provide a fast implementation using Template Model Builder. The key focus is inference with Gaussian random fields under positional uncertainty, and our approach works for both
Gaussian and non-Gaussian likelihoods. A simulation study with a binomial observation model shows that the new approach performs equally or better than the common approach of ignoring jittering, both in terms of more accurate parameter estimates and improved predictive measures. We demonstrate that the improvement would be larger under stronger jittering. An analysis of contraceptive use in Kenya shows that the approach is fast and easy to use in practice.
\end{abstract}


\section{{Introduction}}
Demographic and health indicators are important for monitoring and evaluating progress towards achieving the United Nations' (UN's) Sustainable Development Goals (SDGs) \citep{GA2015}. The DHS program
has collected over 400 surveys in over 90 countries, and surveys are conducted approximately every fifth year in participating countries. DHS surveys primarily use a two-stage cluster sampling design. Georeferenced data are only available based on special permission and a known geographical displacement process is applied before releasing the GPS coordinates of the clusters. Here, DHS aims to balance the risk of disclosure of the respondents while simultaneously preserving useful information for spatial analyses \citep{DHSspatial07}. Urban clusters are displaced up to 2 km, while 99\% of the rural clusters are displaced up to 5 km, and the remaining 1\% up to 10 km. Rural clusters are jittered more to keep the same level of disclosure risk as for urban clusters \citep{vanwey2005}. This approach can be criticized as the actual risk of disclosure is unclear. Alternative procedures exist, which include, for example, location swapping \citep{zhang2017}, space transformations \citep{khoshgozaran2007} and k-anonymity \citep{sweeney2002}. In the field of cyber security so-called strong protection techniques are proposed, see for example \citep{gahi2016}. 

This short report does not assess the quality of the underlying displacement process used by DHS, but proposes a novel and fast geostatistical inference approach to analyse DHS data in the presence of positional uncertainty. 
In a linear geostatistical model with a Gaussian likelihood,
a simple approach to adjust for positional error with a known displacement
distribution 
is to adjust the covariances
between the observed locations, and assume that after marginalising out the 
unknown true locations, the joint distribution is 
still a Gaussian distribution \citep{cressie2003spatial}. However, such
approaches do not easily generalize to generalized linear models. 
\citet{fanshawe2011spatial} describe how to account for positional 
uncertainty in a hierarchical geostatistical model fitted through maximum likelihood estimation for the parameters, but found computational times to be prohibitively slow. 
Later work demonstrates that inference can be made faster through a composite likelihood approach
in the case of a linear geostatistical model with a Gaussian likelihood \citep{fronterre2018geostatistical}.

Recently, \citet{wilson2021estimation} proposed a Bayesian approach for
generalized linear geostatistical models in the context of DHS surveys. Each iteration in their method is composed of two parts. First, Markov chain Monte Carlo (MCMC) is used to sample the true locations,
then the Integrated Nested Laplace Approximations (INLA) method \citep{rue2009approximate} is applied for inference conditional on the true locations. 
This gives an INLA within MCMC approach
\citep{gomez2018markov}, which for each simulation scenario took around 52 hours to run 1,000 iterations on 398 locations.

\citep{warren2016influenceOne} proposed a ``regression calibration (RC)'' method for distance-based analyses that accounts for jittering of DHS clusters by trying to estimate the true distance covariates. They found that the proposed method outperformed the naive method in almost all location and spatial density settings.
In another study, \citep{warren2016influenceTwo} adressed the issue of incorrectly assigning areas to the DHS clusters, when clusters are jittered out of the corresponding true polygons. They proposed a maximum probability covariate (MPC) selection method which allows selecting the most probable covariates. They recommend using MPC to maximize the selection probability of the correct covariates. As a different approach, \citep{DHSspatial11} considered the impact of jittering of DHS clusters from the perspective of spatial interpolation surfaces. They proposed a geostatistical framework for creating interpolated surfaces based on DHS data.

In this short report, we present a novel approach to fit generalized linear geostatistical models that accounts for positional uncertainty in the provided GPS coordinates of the data locations. The key focus is to address the issue of inference under positional uncertantity when modelling spatial variation using Gaussian random fields (GRFs).
Computation time is a key concern and the method needs to be accessible
to analysts without requiring them to write complex code.
 We use a quadrature to integrate out the unknown true
locations so that the likelihood of the observation conditional
on the latent model is a mixture distribution. The random effects
are then integrated out using the Laplace approximation and
automatic differentiation with Template Model Builder \texttt{TMB}, which supports complex, nonlinear latent models with non-Gaussian responses \citep{JSSv070i05}.
Computationally efficient inference is ensured by using the stochastic partial differential 
equation (SPDE) approach \citep{Lindgren:etal:11}, which allows the spatial field to 
be evaluated at any location quickly. We investigate the performance of the new approach compared to standard practice of ignoring jittering in a simulation study focusing on the stability of random effect estimates. 

Increasing populations have a potential to create a huge future demand for the limited resources on food in low- and middle income countries \citep{le2017can, alexandratos2012world}. In order to support family planning policies, as a source of useful insight, we analyse the proportion of contraceptive use among women aged 15-49 based on data from the 2014 Kenya Demographic and Health Survey (KDHS2014) \citep{KDHS2014}. This requires a geostatistical model that can handle a binomial observation model while accounting for jittering.

In Section \ref{sec:dataAndModel}, we describe the KDHS2014 data set and outline the model structure. Section \ref{sec:method} details the proposed method for approximate inference. Section \ref{sec:SimStudy} presents the simulation study and the analysis of contraception use in Kenya.  We end the paper with discussion in Section \ref{sec:discussion}. 
Supplementary results are found in the Supplementary Materials, and all R and C++ code is 
available in the Github repository \href{https://github.com/umut-altay/Supplementary.git}{https://github.com/umut-altay/Supplementary.git}. The repository includes a data statement which outlines the application procedure to download the contraception data from DHS. 

\section{{Data and Model Structure}}

\begin{figure}
\centering
\includegraphics[width=2.85in]{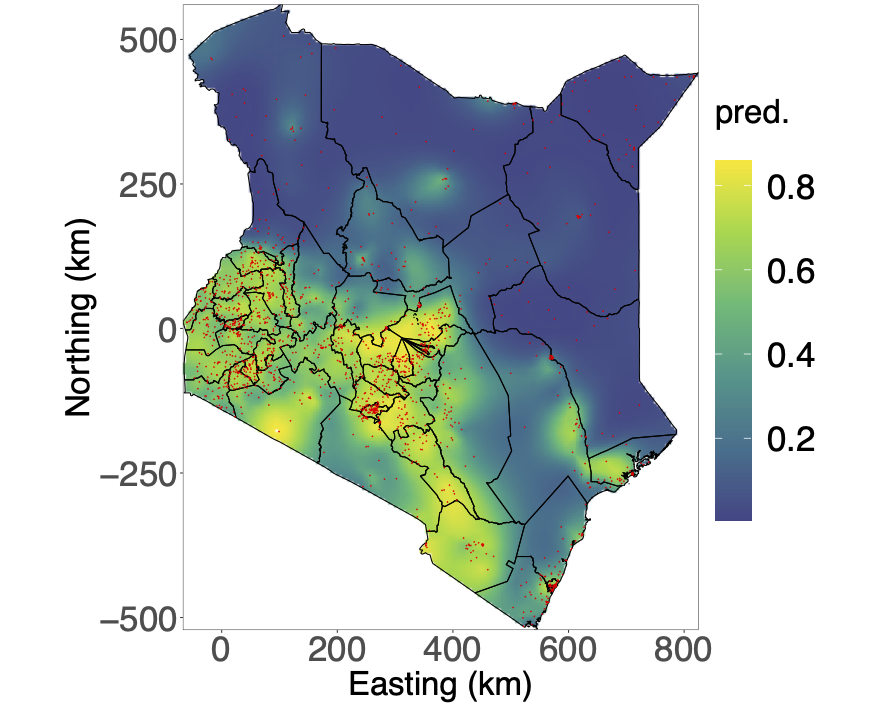} 
\includegraphics[width=2.85in]{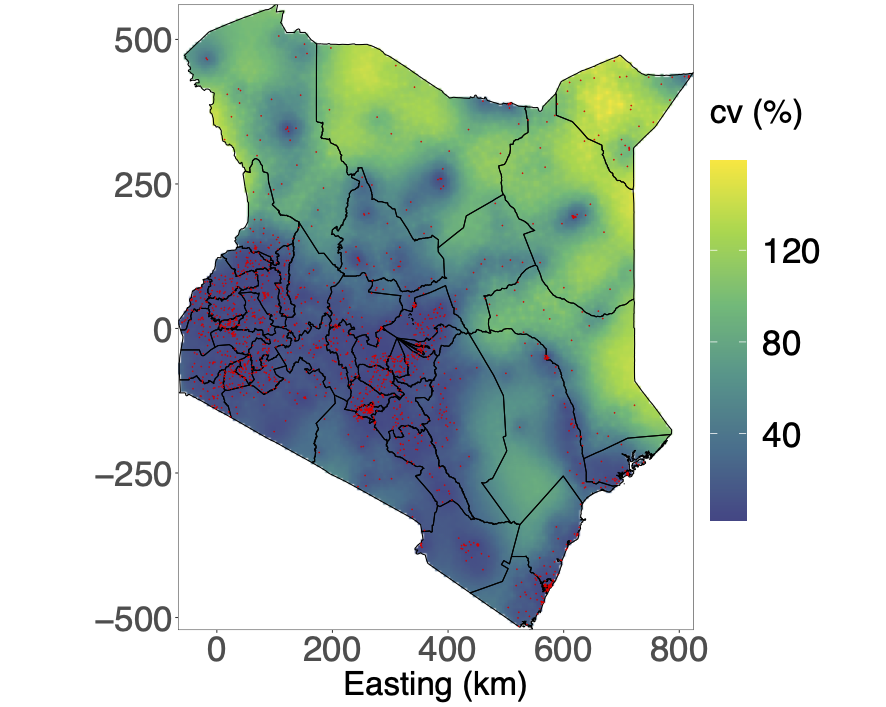}
\caption{Predicted posterior expectations (``pred.'') for the probabilities of using any contraceptive method (left) and the corresponding coefficients of variation (CV) (right) with the new approach. The red points indicate the (jittered) locations of the $C=$1,583 clusters in Kenya.\label{fig:estimates}}
\end{figure}

 Kenya consists of 47 counties, where every location in Kenya is classified as ``urban'' or ``rural''. KDHS2014 contains 1,594 observed clusters,  where the true GPS coordinates have been jittered by the standard DHS procedure restricted so that each GPS location cannot be displaced outside its 
original county.
\label{sec:dataAndModel}
 After eliminating clusters whose coordinates did not match with their designated county or had invalid GPS coordinates, $C = $1,583 clusters
remained. Figure \ref{fig:estimates} shows the geography together with the estimates which are obtained at the end of Section \ref{sec:SimStudy} from the model that we construct to account for jittering  in the observation locations. Similar figures are available in Section 4 of Supplementary Materials, for the standard model that does not account for jittering.

In total, there were 31,079 interviewed women and 17,500 among them reported having used a contraceptive method in 2014. For clusters
$c = 1, \ldots, C$, let $n_c$ denote the number of interviewed women  aged 15--49, $y_c$ the number of those women who
have used contraceptive methods, and $\boldsymbol{s}_c\in\mathbb{R}^2$ the jittered spatial location. The unknown true location is denoted
as $\boldsymbol{s}_c^*\in\mathbb{R}^2$.
We model the risk of contraception use at location $\boldsymbol{s}^*$ as
\[
    \text{logit}(r(\boldsymbol{s}^*)) =  \mu + u(\boldsymbol{s}^*), \quad \boldsymbol{s}^*\in \mathbb{R}^2,
\]
where $\mu$ is an intercept and $u(\cdot)$ is
a GRF with a Mat\'ern covariance function
with marginal variance $\sigma_\mathrm{S}^2$, range $\rho_\mathrm{S}$, 
and fixed smoothness $\nu = 1$. We observe $n_c$ individuals exposed
to this risk and use $y_c | r(\boldsymbol{s}_c^*) \sim \text{Binomial}(n_c, r(\boldsymbol{s}_c^*))$ independently for $c = 1, \ldots, C$.

Under the DHS jittering scheme, for an urban cluster $c$ with a maximum jittering distance of $2\, \mathrm{km}$, 
%which can be jittered up to $2\, \mathrm{km}$,
the jittering distribution is 
\begin{equation}
    \pi_\mathrm{U}(\boldsymbol{s}_c|\boldsymbol{s}_c^*) \propto \frac{\mathbb{I}(A(\boldsymbol{s}_c) =A( \boldsymbol{s}_c^*))\cdot\mathbb{I}( d(\boldsymbol{s}_c,\boldsymbol{s}_c^*)<2)}{ d(\boldsymbol{s}_c,\boldsymbol{s}_c^*)}, \quad \boldsymbol{s}_c \in \mathbb{R}^2,
 \label{eq:obsLoc}
\end{equation}
where, $d(\boldsymbol{s}_c,\boldsymbol{s}_c^*)$ is the distance between $\boldsymbol{s}_c^*$ and $\boldsymbol{s}_c$, $\mathbb{I}$ is an indicator function, and clusters
are jittered independently. Similarly, for a rural cluster $c$, with a maximum jittering distance of $5\, \mathrm{km}$ (and the 1 percent of clusters with a maximum jittering distance of $10\, \mathrm{km}$), the jittering distribution is
\begin{equation}
    \pi_\mathrm{R}(\boldsymbol{s}_c|\boldsymbol{s}_c^*) \propto \frac{\mathbb{I}(A(\boldsymbol{s}_c) =A( \boldsymbol{s}_c^*))}{d(\boldsymbol{s}_c, \boldsymbol{s}_c^*)} \left[\frac{99\mathbb{I}(d(\boldsymbol{s}_c,\boldsymbol{s}_c^*)<5)}{100} + \frac{\mathbb{I}( d(\boldsymbol{s}_c,\boldsymbol{s}_c^*)<10)}{100}\right], \quad \boldsymbol{s}_c \in \mathbb{R}^2.
 \label{eq:obsLoc1}
\end{equation}

The binomial observation model is combined with the location likelihoods in Equations \eqref{eq:obsLoc} and \eqref{eq:obsLoc1}
to give the complete observation model. The underlying latent model is
\[
    \mu \sim \mathcal{N}(0, 1000), \quad (u(\boldsymbol{s}_1^*) \ \ldots \ u(\boldsymbol{s}_c^*))^\mathrm{T}|\sigma_\mathrm{S}^2, \rho_\mathrm{S}, \boldsymbol{s}_1^*, \ldots, \boldsymbol{s}_C^* \sim \mathcal{N}(\boldsymbol{0}, \Sigma), 
\]
where the covariance matrix $\Sigma$ is a function of the unknown
true locations $\boldsymbol{s}_1^*, \ldots, \boldsymbol{s}_C^*$ and
the parameters $\sigma_\mathrm{S}^2$ and $\rho_\mathrm{S}$. We use the
penalised complexity (PC) prior for Mat\'ern GRFs \citep{fuglstad:etal:19a} for $\sigma_\mathrm{S}^2$ and $\rho_\mathrm{S}$ with $\mathrm{P}(\sigma_\mathrm{S} > 1) = 0.05$ and 
$\mathrm{P}(\rho > \rho_0) = 0.50$, i.e., $\rho_0$ is the \emph{a priori} median range.  We use uniform priors for $\boldsymbol{s}_c^*$, this effectively implies that all locations $\boldsymbol{s}_c^*$ such that 
$||\boldsymbol{s}_c-\boldsymbol{s}_c^*||<10$ are
considered equally likely for $c = 1, \ldots, C$.

\section{Approximating the Posterior Under Positional Uncertainty}
\label{sec:method}
The SPDE model decomposes the spatial effect into a linear combination of compactly supported basis functions, $u(\boldsymbol{s}) = \sum_{i=1}^K w_i \phi(s)$, for basis function $\phi_1,\ldots,\phi_K$, and where the $K$-vector of basis weights $\boldsymbol{w} = (w_1 \ \ldots \ w_K)^\mathrm{T}$ follows a multivariate Gaussian distribution with zero mean and with precision matrix set so as to approximate a Mat\'{e}rn covariance structure. This results in a highly sparse precision matrix for the basis weights, and causes the likelihood evaluation to require only $O(K^{3/2}n)$ operations for $K$ basis elements and $n$ observations \citep{Lindgren:etal:11}.

We treat the unknown true locations as nuisance parameters, integrating them out of the likelihood and posterior. Letting $\boldsymbol{\beta} = (\mu)^\mathrm{T}$ be the vector of fixed effect coefficients for the linear predictor $\eta(\cdot)$, the full likelihood can be factorized into a product of likelihoods for individual observations: $\pi(\boldsymbol{y}, \boldsymbol{s}_1, \ldots, \boldsymbol{s}_n \vert \boldsymbol{w}, \boldsymbol{ \beta}, \boldsymbol{ \theta}_L) = \prod_{i=1}^n \pi(y_i, \boldsymbol{s}_i \vert \boldsymbol{w}, \boldsymbol{ \beta}, \boldsymbol{ \theta}_L)$. The likelihood for an individual observation can then be calculated by integrating over the distribution of its possible true spatial locations:
\begin{align}
 \pi(y_i, \boldsymbol{s}_i \vert \boldsymbol{w},  \boldsymbol{\beta} , \boldsymbol{ \theta}_L) &=  \int_{\mathbb{R}^2}  \pi (y_i, \boldsymbol{s}_i \vert \boldsymbol{s}_i^*, \boldsymbol{w}, \boldsymbol{ \beta}, \boldsymbol{ \theta}_L)  \pi (\boldsymbol{s}_i^* \vert \boldsymbol{w}, \boldsymbol{ \beta}, \boldsymbol{ \theta}_L) \ \mathrm{d}\boldsymbol{s}_i^* \nonumber \\
 &= \int_{\mathbb{R}^2} \pi (y_i \vert  \eta(\boldsymbol{s}_i^*), \boldsymbol{ \theta}_L) \pi (\boldsymbol{s}_i \vert \boldsymbol{s}_i^*) \pi (\boldsymbol{s}_i^*) \ \mathrm{d}\boldsymbol{s}_i^*. \label{eq:analyticIndividualIntegral}
\end{align}

Since the integral in \eqref{eq:analyticIndividualIntegral} is two-dimensional, it can be well-approximated for each $i$ via quadrature. We will integrate by selecting a single integration point at $\boldsymbol{s}_i$, and then building more `rings' of points around $\boldsymbol{s}_i$. Let $m_{ij}$ denote the number of integration points for observation $i$ in ring $j$. Each numerical integration point, given by $\boldsymbol{s}_{ijk}^*$ for observation $i$, ring $j$, and index $k = 1,\ldots,m_{ij}$ has an associated integration weight given by $\lambda_{ijk}$. If we assume there are $J^i$ rings in total (counting $\boldsymbol{s}_{i11}^* = \boldsymbol{s}_i$ as the first ring), then we can approximate the integral in \eqref{eq:analyticIndividualIntegral} numerically as follows:
\begin{align}
\int_{\mathbb{R}^2} \pi (y_i \vert  \eta(\boldsymbol{s}_i^*), \boldsymbol{ \theta}_L) \pi (\boldsymbol{s}_i \vert \boldsymbol{s}_i^*) \pi (\boldsymbol{s}_i^*) \ \mathrm{d}\boldsymbol{s}_i^* &= \int \pi (y_i \vert  \eta(\boldsymbol{s}_i^*), \boldsymbol{ \theta}_L) \ \mathrm{d} \left [ \pi (\boldsymbol{s}_i \vert \boldsymbol{s}_i^*) \pi (\boldsymbol{s}_i^*) \right ] \nonumber \\
 &\approx \sum_{j=1}^{J^i} \sum_{k=1}^{m_{ij}} \lambda_{ijk} \pi (y_i \vert  \eta(\boldsymbol{s}_{ijk}^*), \boldsymbol{ \theta}_L), \label{eq:numericalIntegral}
\end{align}
\noindent
where $\lambda_{ijk} \propto \int_{A_{ijk}} \pi (\boldsymbol{s}_i \vert \boldsymbol{s}_i^*) \pi (\boldsymbol{s}_i^*) \ \mathrm{d}\boldsymbol{s}_i^*$, and $A_{ijk}$ is the area associated with integration point $\boldsymbol{s}_{ijk}^*$, and is defined in the Supplement in Section 6. We will take $m_{i1}=1$, and $m_{ij}=15$ for all other $j>1$ so that there are $1 + 15(J^i-1)$ integration points in total. We may assume $\sum_{ij} \sum_k \lambda_{ijk} = 1$ for each $i$, since the scaling of these weights cancels in the posterior. Hence, if $\pi (\boldsymbol{s}_{ijk}^*)$ is constant over the support of $\pi (\boldsymbol{s}_i \vert \boldsymbol{s}_i^*)$, then $\lambda_{ijk} \propto \int_{A_{ijk}} \pi (\boldsymbol{s}_i \vert \boldsymbol{s}_i^*) \ \mathrm{d}\boldsymbol{s}_i^*$. If, however, it is also known that observation $i$ lies in spatial region $R[i]$, and $\pi (\boldsymbol{s}_i \vert \boldsymbol{s}_i^*)$ has any mass outside of $R[i]$, then the weights are: $\lambda_{ijk} \propto \int_{A_{ijk} \cap R[i]} \pi (\boldsymbol{s}_i \vert \boldsymbol{s}_i^*) \ \mathrm{d}\boldsymbol{s}_i^*$.

\begin{figure}
\centering
\includegraphics[width=2.85in]{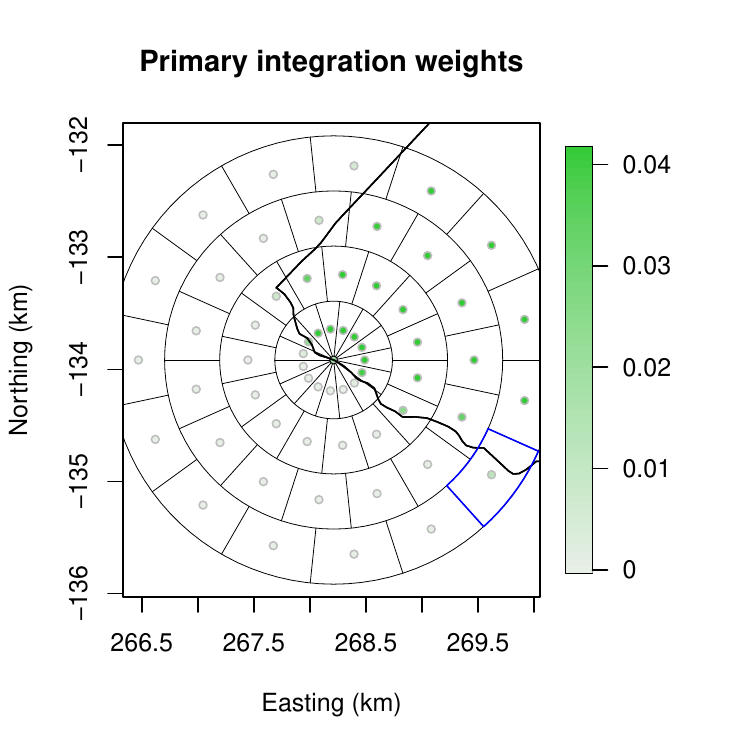} \includegraphics[width=2.85in]{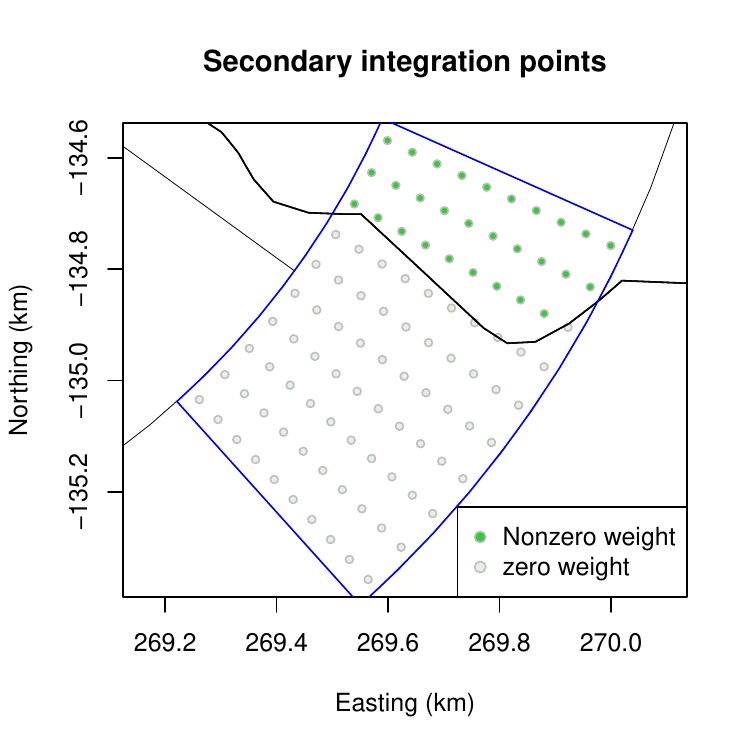}
\caption{Primary integration points and their weights for a cluster in Nairobi (left), and an integration region with associated secondary integration points (right). The integration region is outlined in blue, and Nairobi is outlined in black in both plots.}
\label{fig:integration}
\end{figure}

If observation $i$ is within jittering distance of the boundary of $R[i]$, then its integration weights must be adjusted accordingly. For the $ijk$-th integration region, we approximate $\lambda_{ijk} \propto \int_{A_{ijk} \cap R[i]} \pi (\boldsymbol{s}_i \vert \boldsymbol{s}_i^*) \ \mathrm{d}\boldsymbol{s}_i^*$ numerically by subdividing $A_{ijk}$ into a $10\times10$ grid of `secondary' integration regions, each with an associated secondary integration point at the center of mass of $\pi (\boldsymbol{s}_i \vert \boldsymbol{s}_i^*)$ on that secondary integration region. We calculate the center of mass radius by shrinking the midpoint radial coordinate of the secondary integration regions within the subregions by an equivalent factor as in Equation 2 of Section 6 in the Supplementary Material, except replacing the subregion boundary angles $a_{ij2} - a_{ij1}$ (defined in the Supplement in Section 6) with $(a_{ij2} - a_{ij1})/10$. We then scale $\lambda_{ijk}$ depending on the proportion of associated subintegration points in $R[i]$.  This is equivalent to assuming that all secondary integration points associated with a given integration region have approximately equal weight. This adjustment to the weights, as well as the integration regions and points for an urban cluster in Nairobi, are depicted in Figure \ref{fig:integration}. Technical details regarding the generation of the integration points, weights, and regions, including derivations, are given in Section 6 of the Supplementary Material.

We implement the above model in C++ using TMB, which integrates out $\boldsymbol{w}$, and uses autodifferentiation to maximize and takes a Laplace approximation of the posterior. As a result, the proposed method has the computational advantages of both the SPDE model and of its implementation in TMB. If $M_{\tiny \mbox{P}}$ and $M_{\tiny \mbox{S}}$ are respectively the average number of primary integration points per observation and the number of secondary integration points per primary integration point, then our method still only requires $O(M_{\tiny \mbox{P}} M_{\tiny \mbox{S}} n K^{3/2}) = O(n K^{3/2})$ computational operations per likelihood evaluation. The autodifferentiation of TMB also helps to reduce the number of operations required for optimizing the approximated posterior.

The integration weights before correction for boundary effects, the radial displacement of the integration points, and the number of points per integration ring are given in Table 5 in Section 6 of the  Supplementary Material.

%%%%%%%%%% {
\begin{comment}
\begin{figure}
\centering
\includegraphics[width=3.2in]{jitteringPriorRadial_numPointsUrban11_numPointsRural16.pdf} 
\caption{Numerical integration weight and jittering process density as a function of radial displacement.}
\label{fig:integrationRadial}
\end{figure}
\end{comment}
%%%%%%%%%% }

\section{Simulation Study and Analysis of Contraception Use in Kenya}
\label{sec:SimStudy}

We evaluate the gain when accounting for jittering through a simulation study where
data is generated according to the model described in Section \ref{sec:dataAndModel}.
The GRF is simulated using marginal variance 
$\sigma_\mathrm{S}^2 = 1$, spatial ranges $\rho_\mathrm{S}\in \{160, 340\}$ (km), and smoothness $\nu = 1$.
These ranges correspond to approximately $1/5$ and $2/5$ of the extent of Kenya in West-East-direction. We fix the true coordinates to match the $C = 1,583$ clusters with
reliable location information in KDHS2014, and set the intercept $\mu = 0$, which corresponds to 50\% contraception use. This is motivated by the fact that contraception use in Kenya has strong
spatial variation, but with a national level around 58\% \citep{KDHS2014}. Datasets
are generated by simulating $y_c$ at
location $\boldsymbol{s}_c^*$ from a binomial distribution where the success probability is 
$r(\boldsymbol{s}_c^*)$ and the number of trials $n_c = 100$ for $c = 1, \ldots, C$.
Section 3 in the Supplementary Material presents the corresponding study 
with a Gaussian observation model.

For each of the two ranges, we simulate the GRF and responses repeatedly to give
$50$ datasets. To each of these datasets we apply two jittering strategies: 1) standard DHS jittering, and 2) DHS jittering with maximum distances multiplied with 4 (termed 4 $\times$ DHS jittering). This gives 200 datasets for the four combinations of ranges (160 km and 340 km) and jittering options. For each dataset we fit a standard spatial model that assumes locations
are correct (Model-S) and the new model that accounts for positional uncertainty (Model-J). For the model specification in Section \ref{sec:dataAndModel}, we set the \emph{a priori} median
of range $\rho_0$ equal to true range. After fitting
the model, we compute the continuous rank probability score (CRPS) and the logarithmic score (log-score) \citep{gneiting2007strictly} for 1,000 evenly distributed prediction locations (shown in  Figure 1 in the Supplementary Material).

Posterior inference is approximately Bayesian using TMB, and
parameter estimates are computed using posterior medians.
Table \ref{tab:simStudy} shows that there is less bias in the parameter estimates
when using
Model-J than Model-S. The difference between the two approaches becomes larger
for 4 $\times$ DHS jittering than standard DHS jittering. The positional uncertainty
in Model-J gives larger $95\%$ credible intervals (CIs) for the parameters compared to the Model-S, and the difference is larger for more jittering.

\begin{figure}
  \begin{subfigure}[t]{.5\textwidth}
    \centering
    \includegraphics[width=\linewidth]{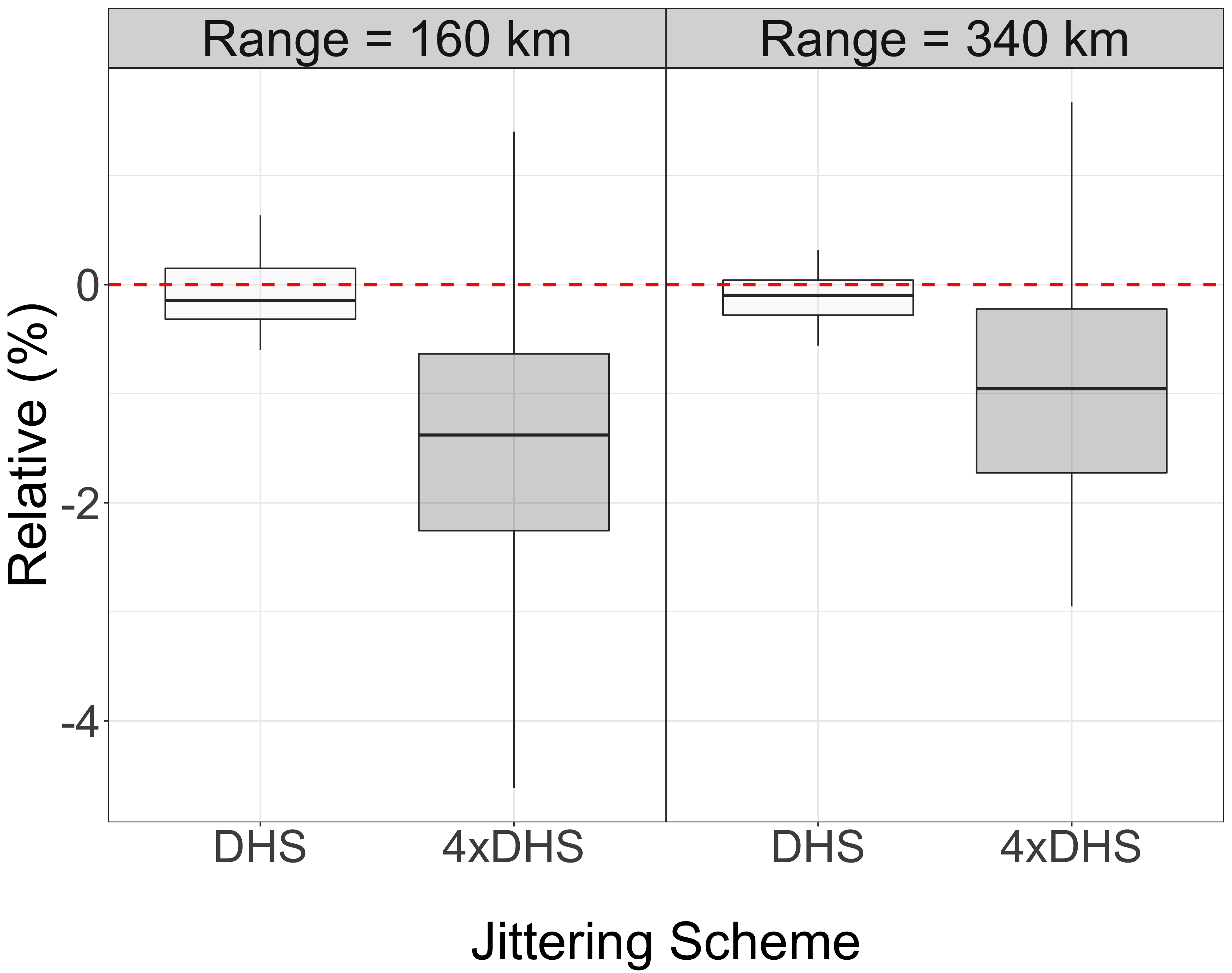}
    \caption{Pair-wise relative differences $\left (100 \times \frac{{CRPS}_{J} - {CRPS}_{S}}{{CRPS}_{S}} \right )$ in CRPS\label{fig:scoreCRPS}}
  \end{subfigure}
  \hfill
  \begin{subfigure}[t]{.5\textwidth}
    \centering
    \includegraphics[width=\linewidth]{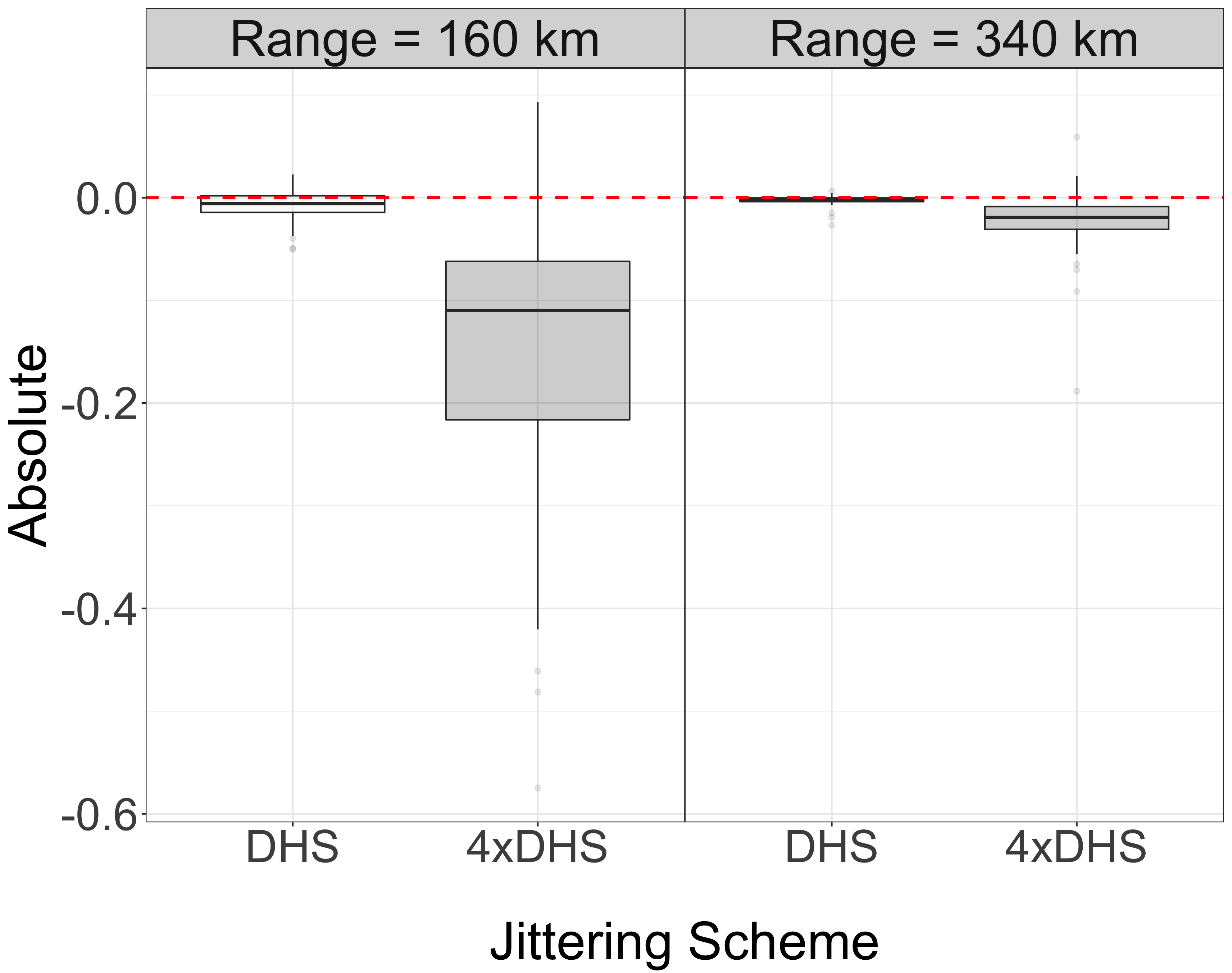}
    \caption{Pair-wise absolute differences in Log Scores\label{fig:scoreLog}}
  \end{subfigure}
  \caption{Pair-wise differences in CRPS and Log Scores that are obtained from Model-J and Model-S for binomial observation model.\label{fig:score}}
\end{figure}

Figure \ref{fig:scoreCRPS} shows a minor improvement in relative difference in CRPS for the prediction locations with Model-J compared
to Model-S under standard DHS jittering. For 4 $\times$ DHS jittering, there is a clear 
improvement. Figure \ref{fig:scoreLog} shows similar behavior for the log-score,
but with a less clear difference with $\rho_\mathrm{S} = 340$ km and 4 $\times$ DHS jittering.
There were only minor differences in the average
coverage of the predictive distributions as shown in Table 3 in the Supplementary Material.
A corresponding simulation study with a Gaussian observation model in Section 3 in the Supplementary Materials leads to similar conclusions, and demonstrates that a nugget variance is overestimated when jittering is not accounted for.

\begin{table}
\centering
    \caption{Average biases and average 95\% CI lengths of parameter estimates under Model-J. We use absolute bias for $\mu$ and relative bias for $\rho_\mathrm{S}$ and $\sigma_\mathrm{S}^2$. The corresponding values
    using Model-S are shown in parentheses.\label{tab:simStudy}}
    \begin{tabular}{ll|ll|ll|}
         \multicolumn{2}{c}{}  &
        \multicolumn{2}{c}{\textbf{DHS jittering}} & \multicolumn{2}{c}{\textbf{4xDHS jittering}}\\
        \textbf{Parameter} & \textbf{Truth}
       & \textbf{Bias} & \textbf{CI length} & \textbf{Bias} & \textbf{CI length} \\
       \hline
        \multicolumn{2}{l}{\textbf{Short range}} & \multicolumn{2}{l}{} & \multicolumn{2}{l}{} \\
        ~~$\mu$ & 0 & -0.03 \emph{(-0.03)} & 0.79 \emph{(0.77)} & -0.03 \emph{(-0.03)} & 0.84 \emph{(0.69)} \\ 
        ~~$\rho_\mathrm{S}$      & 160 & -3\% \emph{(-6\%)} & 69 \emph{(65)} & 7\% \emph{(-13\%)} & 82 \emph{(59)} \\ 
        ~~$\sigma^{2}_\mathrm{S}$ & 1 & -2\% \emph{(-2\%)} & 0.34 \emph{(0.33)} & -4\% \emph{(-7\%)} & 0.37 \emph{(0.30)} \\ 
        \multicolumn{2}{l}{\textbf{Long range}} & & & \\
        ~~$\mu$ & 0 & -0.04 \emph{(-0.04)} & 1.24 \emph{(1.23)} & -0.10 \emph{(-0.10)} & 1.27 \emph{(1.18)} \\ 
        ~~$\rho_\mathrm{S}$      & 340 & -7\% \emph{(-9\%)} & 203 \emph{(200)} & -3\% \emph{(-10\%)} & 221 \emph{(199)} \\ 
        ~~$\sigma^{2}_\mathrm{S}$ & 1 & -7\% \emph{(-8\%)} & 0.51 \emph{(0.50)} & -9\% \emph{(-11\%)} & 0.52 \emph{(0.48)} \\  
\end{tabular}
\end{table}

We apply the new approach to the contraception use dataset described in Section \ref{sec:dataAndModel}. Model-S and Model-J
were estimated in 21 seconds and 8 minutes, respectively. For the real data analysis we again place a PC prior on the spatial range parameter, setting the median spatial range to $\rho_0 = 160$ km. The estimated contraception use probabilities and coefficients of variation for Model-J are shown in Figure \ref{fig:estimates}. The map
shows contraception use is high in the southwest
direction and low in northeast. 
On average CVs are 2.7\% higher for Model-J relative
to Model-S and point estimates are nearly indistinguishable; see
Section 5 of the Supplementary Materials for more details and
figures for Model-S.

\section{Discussions and Conclusions}
\label{sec:discussion}
Our simulation study suggests that
accounting for the presence of jittering, or positional uncertainty, in the geostatistical analysis of DHS data on contraception use leads to more accurate parameter estimates than a standard geostatistical analysis. The improvement becomes more pronounced if more jittering is applied than DHS applies by default. Further, we see slight improvement in predictive quality under standard DHS jittering, and
this improvement becomes clearer for higher amounts of jittering. Our novel approach represents a major improvement over
existing inference approaches that are suitable for binomial observation models such as INLA within MCMC, where computation time is measured in days \citep{wilson2021estimation}. The computation time of the new approach is measured in minutes as compared to days for INLA within MCMC.

In the simulation study we encountered
numerical issues when fitting a small number of the simulations. These occurred when the amount of jittering was large compared to the spatial range. In the case of range 160 km and 4$\times$DHS jittering, 2 out of 50 model runs crashed. Though, this amount of jittering
is large compared to what is used in practice by DHS,
but there is a need for future investigation into
methods that are more stable for higher amounts of positional uncertainty. 
The focus of this paper is to present a fast geostatistical model that accounts for jittering during inference. Our approach supports generalized linear geostatistical models with a wide variety of non-Gaussian observation models due to its implementation in TMB. It also is applicable in the context of other known jittering distributions, such as in cases where the administrative area of a cluster is known, but the exact location within the area is not. One limitation, however, is that the computational efficiency will decrease when large displacements
of coordinates are possible relative to the size of the domain
of interest. This is due to decreasing sparsity in the precision matrix induced by jittering distributions overlapping with more spatial basis functions. An interesting potential direction of future research would be to model positional uncertainty when including spatially varying covariates. Furthermore, it would be interesting to investigate the accuracy of the approach presented in this paper to other jittering strategies such as swapping and truncating
can also be applied \citep{DHSspatial07}.

\bibliography{mybibfile}

\begin{thebibliography}{}

\bibitem[Alexandratos and Bruinsma, 2012]{alexandratos2012world}
Alexandratos, N. and Bruinsma, J. (2012).
\newblock World agriculture towards 2030/2050: the 2012 revision.

\bibitem[Burgert et~al., 2013]{DHSspatial07}
Burgert, C.~R., Colston, J., Roy, T., and Zachary, B. (2013).
\newblock Geographic displacement procedure and georeferenced datarelease
  policy for the {D}emographic and {H}ealth {S}urveys.
\newblock \url{https://dhsprogram.com/pubs/pdf/SAR7/SAR7.pdf}.
\newblock DHS Spatial Analysis Reports No. 7.

\bibitem[Cressie and Kornak, 2003]{cressie2003spatial}
Cressie, N. and Kornak, J. (2003).
\newblock Spatial statistics in the presence of location error with an
  application to remote sensing of the environment.
\newblock {\em Statistical {S}cience}, pages 436--456.

\bibitem[Fanshawe and Diggle, 2011]{fanshawe2011spatial}
Fanshawe, T. and Diggle, P. (2011).
\newblock Spatial prediction in the presence of positional error.
\newblock {\em Environmetrics}, 22(2):109--122.

\bibitem[Fronterr{\`e} et~al., 2018]{fronterre2018geostatistical}
Fronterr{\`e}, C., Giorgi, E., and Diggle, P. (2018).
\newblock Geostatistical inference in the presence of geomasking: a
  composite-likelihood approach.
\newblock {\em Spatial {S}tatistics}, 28:319--330.

\bibitem[Fuglstad et~al., 2019]{fuglstad:etal:19a}
Fuglstad, G.-A., Simpson, D., Lindgren, F., and Rue, H. (2019).
\newblock Constructing priors that penalize the complexity of {G}aussian random
  fields.
\newblock {\em Journal of the American Statistical Association}, 114:445--452.

\bibitem[Gahi et~al., 2016]{gahi2016}
Gahi, Y., Guennoun, M., and Mouftah, H.~T. (2016).
\newblock Big data analytics: Security and privacy challenges.
\newblock In {\em 2016 IEEE Symposium on Computers and Communication (ISCC)},
  pages 952--957. IEEE.

\bibitem[{General Assembly of the United Nations}, 2015]{GA2015}
{General Assembly of the United Nations} (2015).
\newblock Resolution adopted by the {G}eneral {A}ssembly on 25 {S}eptember
  2015.
\newblock A/RES/70/1.

\bibitem[Gething et~al., 2013]{DHSspatial11}
Gething, P., Tatem, A., Bird, T., and Burgert-Brucker, C.~R. (2013).
\newblock Creating spatial interpolation surfaces with {DHS} data.
\newblock \url{https://dhsprogram.com/pubs/pdf/SAR11/SAR11.pdf}.
\newblock DHS Spatial Analysis Reports No. 11.

\bibitem[Gneiting and Raftery, 2007]{gneiting2007strictly}
Gneiting, T. and Raftery, A.~E. (2007).
\newblock Strictly proper scoring rules, prediction, and estimation.
\newblock {\em {Journal of the American Statistical Association}},
  102(477):359--378.

\bibitem[G{\'o}mez-Rubio and Rue, 2018]{gomez2018markov}
G{\'o}mez-Rubio, V. and Rue, H. (2018).
\newblock Markov chain {M}onte {C}arlo with the integrated nested {L}aplace
  approximation.
\newblock {\em Statistics and {C}omputing}, 28(5):1033--1051.

\bibitem[Khoshgozaran and Shahabi, 2007]{khoshgozaran2007}
Khoshgozaran, A. and Shahabi, C. (2007).
\newblock Blind evaluation of nearest neighbor queries using space
  transformation to preserve location privacy.
\newblock In {\em International symposium on spatial and temporal databases},
  pages 239--257. Springer.

\bibitem[Kristensen et~al., 2016]{JSSv070i05}
Kristensen, K., Nielsen, A., Berg, C.~W., Skaug, H., and Bell, B.~M. (2016).
\newblock {TMB}: Automatic {D}ifferentiation and {L}aplace {A}pproximation.
\newblock {\em Journal of {S}tatistical {S}oftware}, 70(5):1–21.

\bibitem[Le~Mou{\"e}l and Forslund, 2017]{le2017can}
Le~Mou{\"e}l, C. and Forslund, A. (2017).
\newblock How can we feed the world in 2050? a review of the responses from
  global scenario studies.
\newblock {\em {European Review of Agricultural Economics}}, 44(4):541--591.

\bibitem[Lindgren et~al., 2011]{Lindgren:etal:11}
Lindgren, F., Rue, H., and Lindstr\"{o}m, J. (2011).
\newblock An explicit link between {G}aussian fields and {G}aussian {M}arkov
  random fields: the stochastic differential equation approach (with
  discussion).
\newblock {\em Journal of the Royal Statistical Society, Series B},
  73:423--498.

\bibitem[{National Bureau of Statistics-Kenya} and {ICF International},
  2015]{KDHS2014}
{National Bureau of Statistics-Kenya} and {ICF International} (2015).
\newblock 2014 {KDHS} key findings.
\newblock \url{https://www.dhsprogram.com/pubs/pdf/sr227/sr227.pdf}.

\bibitem[Rue et~al., 2009]{rue2009approximate}
Rue, H., Martino, S., and Chopin, N. (2009).
\newblock Approximate {B}ayesian inference for latent {G}aussian models by
  using integrated nested {L}aplace approximations.
\newblock {\em Journal of the {R}oyal {S}tatistical {S}ociety: Series {B}
  ({S}statistical {M}ethodology)}, 71(2):319--392.

\bibitem[Sweeney, 2002]{sweeney2002}
Sweeney, L. (2002).
\newblock k-anonymity: A model for protecting privacy.
\newblock {\em International journal of uncertainty, fuzziness and
  knowledge-based systems}, 10(05):557--570.

\bibitem[VanWey et~al., 2005]{vanwey2005}
VanWey, L.~K., Rindfuss, R.~R., Gutmann, M.~P., Entwisle, B., and Balk, D.~L.
  (2005).
\newblock Confidentiality and spatially explicit data: Concerns and challenges.
\newblock {\em Proceedings of the National Academy of Sciences},
  102(43):15337--15342.

\bibitem[Warren et~al., 2016a]{warren2016influenceOne}
Warren, J.~L., Perez-Heydrich, C., Burgert, C.~R., and Emch, M.~E. (2016a).
\newblock Influence of demographic and health survey point displacements on
  distance-based analyses.
\newblock {\em Spatial {D}emography}, 4(2):155--173.

\bibitem[Warren et~al., 2016b]{warren2016influenceTwo}
Warren, J.~L., Perez-Heydrich, C., Burgert, C.~R., and Emch, M.~E. (2016b).
\newblock Influence of demographic and health survey point displacements on
  point-in-polygon analyses.
\newblock {\em Spatial {D}emography}, 4(2):117--133.

\bibitem[Wilson and Wakefield, 2021]{wilson2021estimation}
Wilson, K. and Wakefield, J. (2021).
\newblock Estimation of health and demographic indicators with incomplete
  geographic information.
\newblock {\em Spatial and Spatio-temporal Epidemiology}, 37:100421.

\bibitem[Zhang et~al., 2017]{zhang2017}
Zhang, S., Freundschuh, S.~M., Lenzer, K., and Zandbergen, P.~A. (2017).
\newblock The location swapping method for geomasking.
\newblock {\em Cartography and Geographic Information Science}, 44(1):22--34.

\end{thebibliography}

\begin{appendices}

\begin{figure}
    \centering
    \includegraphics[width=\linewidth]{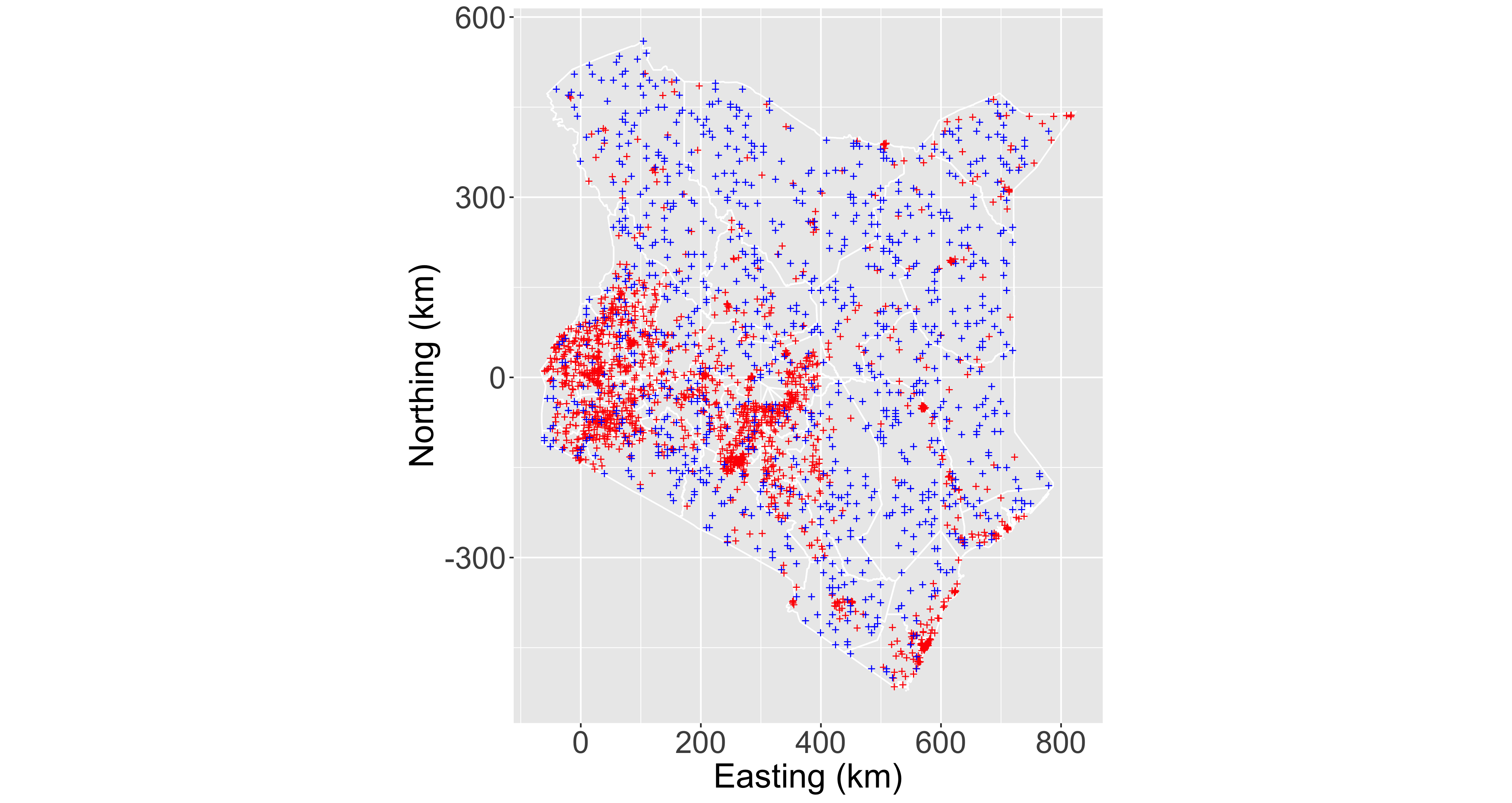}
    \caption{Locations that are used for the study, within Kenya. Jittered locations of the
C = 1,583 clusters are indicated by red. Prediction locations are indicated by blue.\label{tab:locs}}
\end{figure}

This document consists of the supplementary results and materials for our paper titled "Fast geostatistical inference under positional uncertainty: Analysing DHS household survey data". We used (jittered) 2014 Kenya Demographic and Health Survey (KDHS2014) clusters for our study. Figure \ref{tab:locs} shows them together with the prediction locations. The rest of the document is structured as follows:

Section \ref{sec:binomial} presents the supplementary figures of continuous rank probability score (CRPS) and log-score that are obtained from the simulations with the binomial observation model. 

Section \ref{sec:Gaussian} consists of figures and tables of CRPS and log-score that are obtained from the simulations with the Gaussian observation model. Section \ref{sec:coverage} presents the tables of coverage values that are obtained from the simulations with both the binomial and Gaussian observation models. Average computation (model estimation) times that are measured for Model-J during the simulation study under different scenarios are also shared in this section. Section \ref{sec:KDHS2014} shows the results of additional predictions that are done using the binomial model on KDHS2014 contraceptive usage data. Section \ref{sec:numInt} explains how numerical integrations are conducted in our approach. 

\section{Supplementary Results for Binomial Likelihood\label{sec:binomial}}
\begin{figure}
\centering
\includegraphics[width=6in]{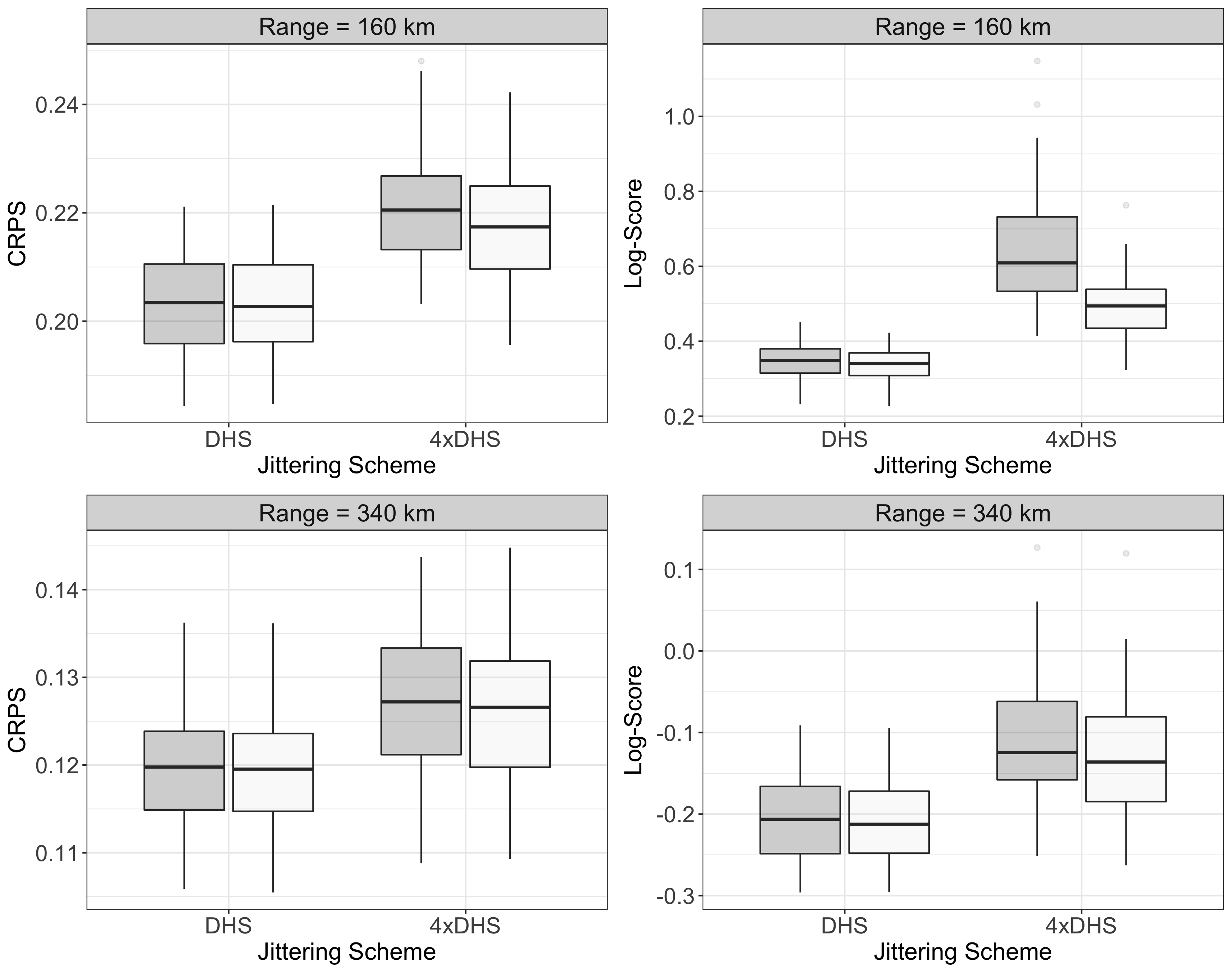} 
\caption{Box-plots of CRPS and log-score values that are obtained from Model-S (boxes with the darker color) and Model-J (boxes with the lighter color) at 1000 prediction locations, for the simulations with the binomial observation model. \label{fig:boxesBinomTRUE}}
\end{figure}

This section presents the supplementary results for the simulation study with the binomial observation model. Figure \ref{fig:boxesBinomTRUE}  shows the box-plots of CRPS and log-score values that are obtained from Model-S and Model-J for the scenarios combining ranges $\rho_\mathrm{S}\in \{160, 340\}$ (km) with jittering schemes (DHS and 4xDHS). Smaller CRPS and log-scores indicate better predictions. Figure \ref{fig:boxesBinomTRUE} shows that Model-J tends to achieve smaller prediction scores and to make better predictions than Model-S as the jittering gets larger. Both models react to the increasing spatial range by providing better predictions.

\section{Simulation Study for Gaussian Likelihood\label{sec:Gaussian}}

\begin{figure}
  \begin{subfigure}[t]{.5\textwidth}
    \centering
    \includegraphics[width=\linewidth]{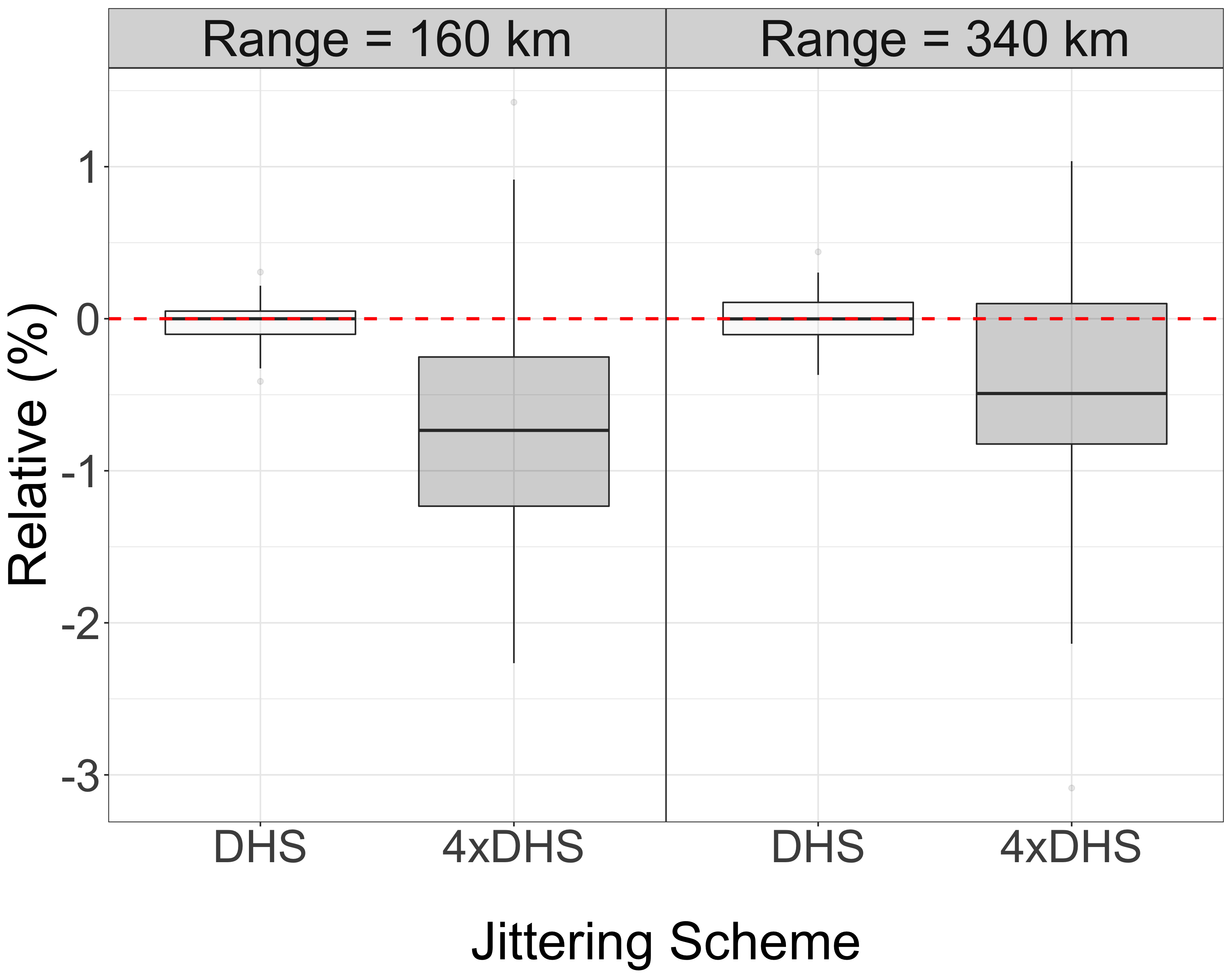}
    \caption{Relative differences 
    ($100*\frac{{CRPS}_{J} - {CRPS}_{S}}{{CRPS}_{S}}$) in CRPS \label{fig:relativeCRPSGaussTRUE}}
  \end{subfigure}
  \hfill
  \begin{subfigure}[t]{.5\textwidth}
    \centering
    \includegraphics[width=\linewidth]{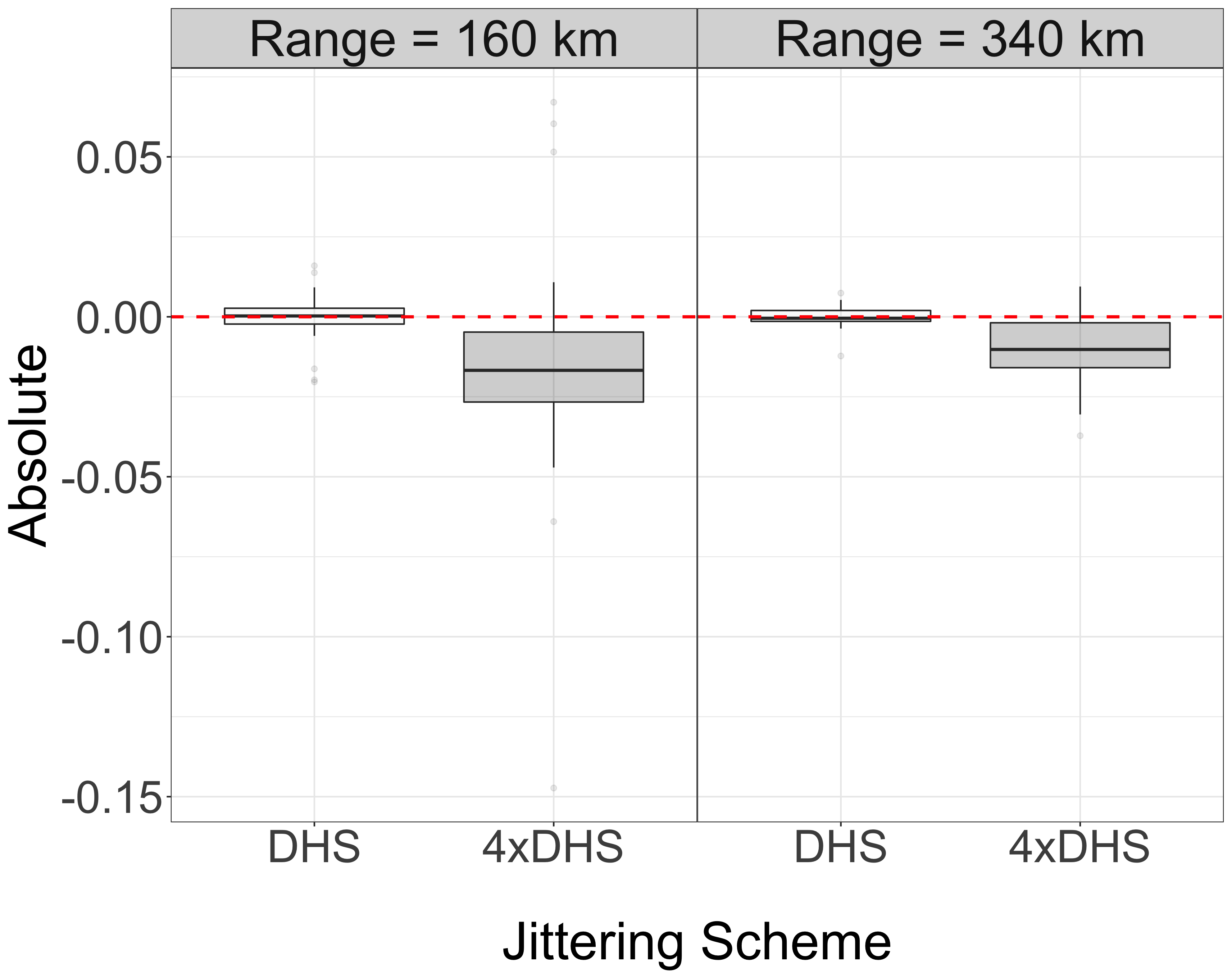}
    \caption{Absolute differences in Log Scores\label{fig:absoluteLogGaussTRUE}}
  \end{subfigure}
  \caption{Box plots of pair-wise differences of the prediction scores that are obtained from Model-S and Model-J at 1000 prediction locations, for the simulations with the Gaussian observation model. \label{fig:diffGaussTRUE}}
\end{figure}

\begin{figure}
\centering
\includegraphics[width=6in]{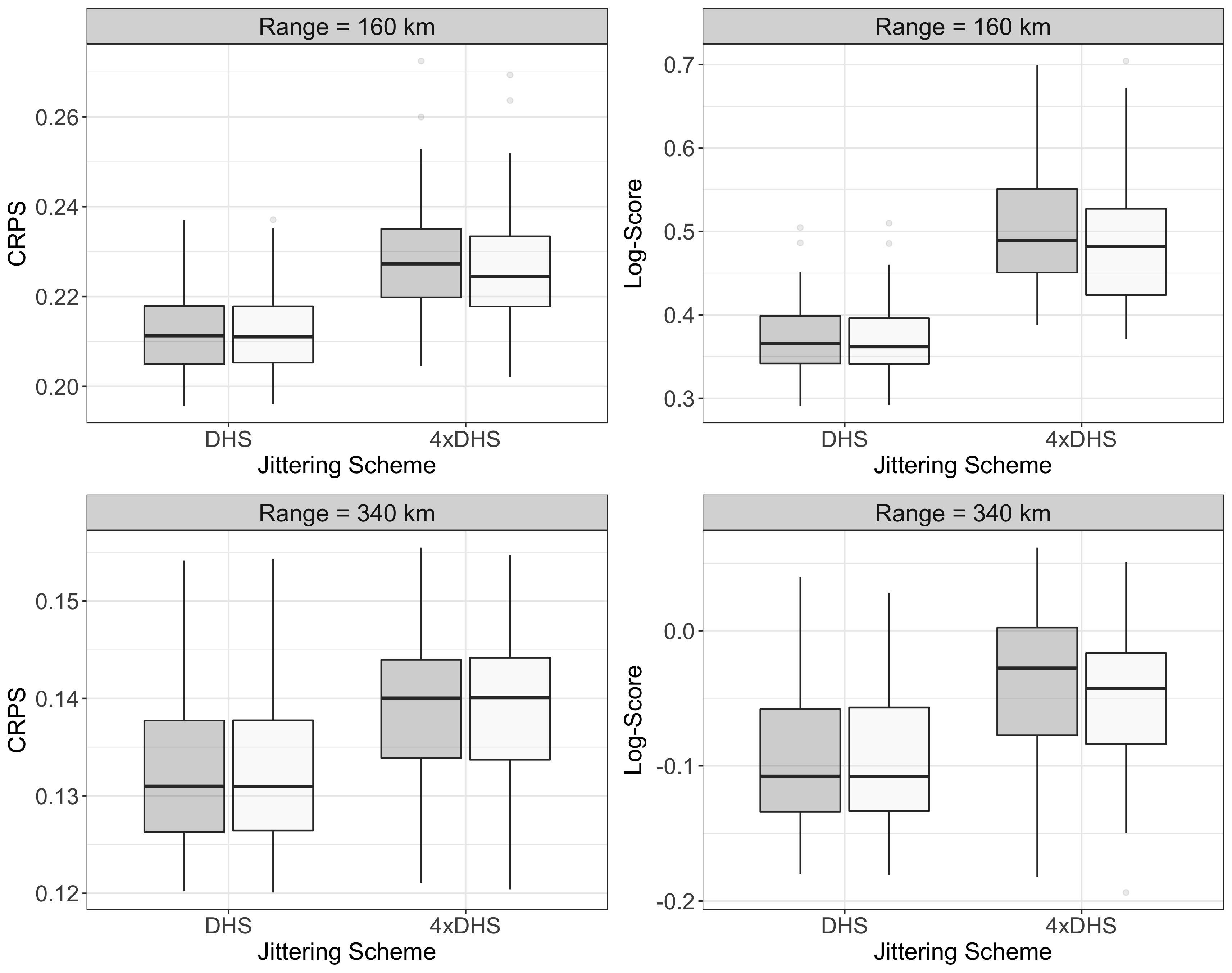} 
\caption{Box-plots of CRPS and log-scores that are obtained from Model-S (boxes with the darker color) and Model-J (boxes with the lighter color) at 1000 prediction locations, for the simulations with the Gaussian observation model. \label{fig:boxesGaussTRUE}}
\end{figure}

This section presents the results of the simulation study with the Gaussian observation model. Figure \ref{fig:diffGaussTRUE} shows the box plots of pair-wise relative differences in CRPS and the absolute differences in log-score for the
prediction locations, with Model-J compared to Model-S. Figure \ref{fig:boxesGaussTRUE} shows the box-plots of CRPS and log-score values for the scenarios combining ranges $\rho_\mathrm{S}\in \{160, 340\}$ (km) and jittering schemes (DHS and 4xDHS). Table \ref{tab:GaussTRUE} presents the average biases and average CI lengths of parameter estimates.

\begin{table}
\centering
    \caption{Average biases and average 95\% CI lengths of parameter estimates. We use absolute bias for $\mu$ and relative bias for $\rho_\mathrm{S}$, $\sigma^{2}_{N}$ and $\sigma_\mathrm{S}^2$. The corresponding values
    using Model-S are shown in parantheses.\label{tab:GaussTRUE}}
    \begin{tabular}{ll|ll|ll|}
         \multicolumn{2}{c}{}  &
        \multicolumn{2}{c}{\textbf{DHS jittering}} & \multicolumn{2}{c}{\textbf{4xDHS jittering}}\\
        \textbf{Parameter} & \textbf{Truth}
       & \textbf{Bias} & \textbf{CI length} & \textbf{Bias} & \textbf{CI length} \\
       \hline
        \multicolumn{2}{l}{\textbf{Short range}} & \multicolumn{2}{l}{} & \multicolumn{2}{l}{} \\
        ~~$\mu$ & 0 & -0.03 \emph{(-0.03)} & 0.80 \emph{(0.79)} & -0.04 \emph{(-0.04)} & 0.86 \emph{(0.84)} \\ 
        ~~$\rho_\mathrm{S}$ & 160 & -0.6\% \emph{(-1\%)} & 77 \emph{(76)} & 8\% \emph{(7\%)} & 89 \emph{(89)} \\ 
        ~~$\sigma^{2}_{N}$ & 0.1 & 6\% \emph{(8\%)} & 0.01 \emph{(0.01)} & 11\% \emph{(38\%)} & 0.02 \emph{(0.02)} \\
        ~~$\sigma^{2}_\mathrm{S}$ & 1 & -3\% \emph{(-3\%)} & 0.35 \emph{(0.35)} & -3\% \emph{(-4\%)} & 0.37 \emph{(0.37)} \\ 
        \multicolumn{2}{l}{\textbf{Long range}} & & & \\
        ~~$\mu$ & 0 & -0.04 \emph{(-0.04)} & 1.25 \emph{(1.25)} & -0.04 \emph{(-0.04)} & 1.23 \emph{(1.23)} \\ 
        ~~$\rho_\mathrm{S}$      & 340 & -7\% \emph{(-7\%)} & 216 \emph{(215)} & -7\% \emph{(-6\%)} & 220 \emph{(222)} \\ 
        ~~$\sigma^{2}_{N}$ & 0.1 & 0.7\% \emph{(1\%)} & 0.01 \emph{(0.01)} & 2\% \emph{(11\%)} & 0.01 \emph{(0.01)} \\
        ~~$\sigma^{2}_\mathrm{S}$ & 1 & -8\% \emph{(-8\%)} & 0.51 \emph{(0.51)} & -10\% \emph{(-10\%)} & 0.50 \emph{(0.50)} \\  
\end{tabular}
\end{table}

\section{Model Estimation Times and Coverage\label{sec:coverage}}
Table \ref{tab:simStudy} shows the average model estimation times (in minutes) obtained by running Model-J on different simulation scenarios. Table \ref{tab:CoverageTRUE} shows the coverage values obtained from each scenario, using both Model-S and Model-J. 

\begin{table}
\centering
    \caption{Average model estimation times with Model-J (in minutes) during the simulation study . \label{tab:simStudy}}
    \begin{tabular}{l|ll|ll|}
        & \multicolumn{2}{c}{\textbf{Short range}} & \multicolumn{2}{c}{\textbf{Long range}}\\
       \textbf{Simulations} & \textbf{DHS jittering} & \textbf{4xDHS jittering} & \textbf{DHS jittering} & \textbf{4xDHS jittering} \\
       \hline
        Binomial & 4.61 & 9.58 & 4.60 & 8.32 \\
        Gaussian & 4.04 & 7.32 & 3.34 & 6.02 \\
\end{tabular}
\end{table}

\begin{table}
\centering
    \caption{Coverage values of Model-J. The corresponding values
    using Model-S are shown in the parantheses.\label{tab:CoverageTRUE}}
    \begin{tabular}{|l|l|l|l|}
    \hline
        \textbf{Simulations} & \textbf{Range} & \textbf{DHS jittering} &\textbf{4xDHS jittering}\\
       \hline
        \textbf{Gaussian} & Short & 0.92 \emph{(0.92)} & 0.89 \emph{(0.89)}\\
        & Long   &0.93 \emph{(0.93)} &0.92 \emph{(0.92)}\\
        \hline
        \textbf{Binomial} & Short &0.91 \emph{(0.91)} &0.87 \emph{(0.88)}\\
        & Long   & 0.93 \emph{(0.93)}&0.91 \emph{(0.90)}\\
        \hline
\end{tabular}
\end{table}

\section{Additional KDHS2014 Contraceptive Usage Results\label{sec:KDHS2014}}
 
\begin{figure}
\centering
\includegraphics[width=2.85in]{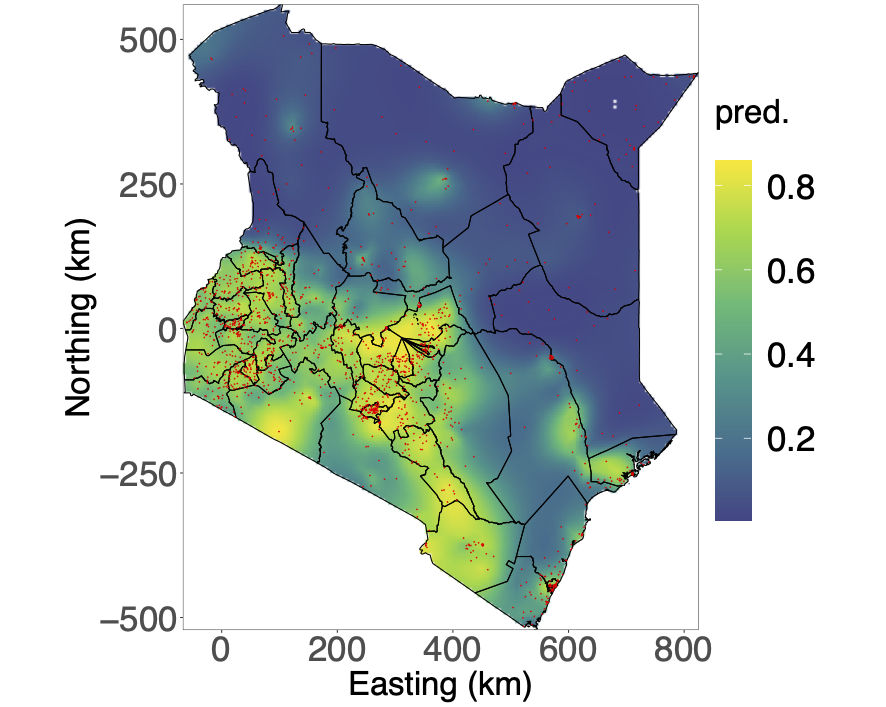} 
\includegraphics[width=2.85in]{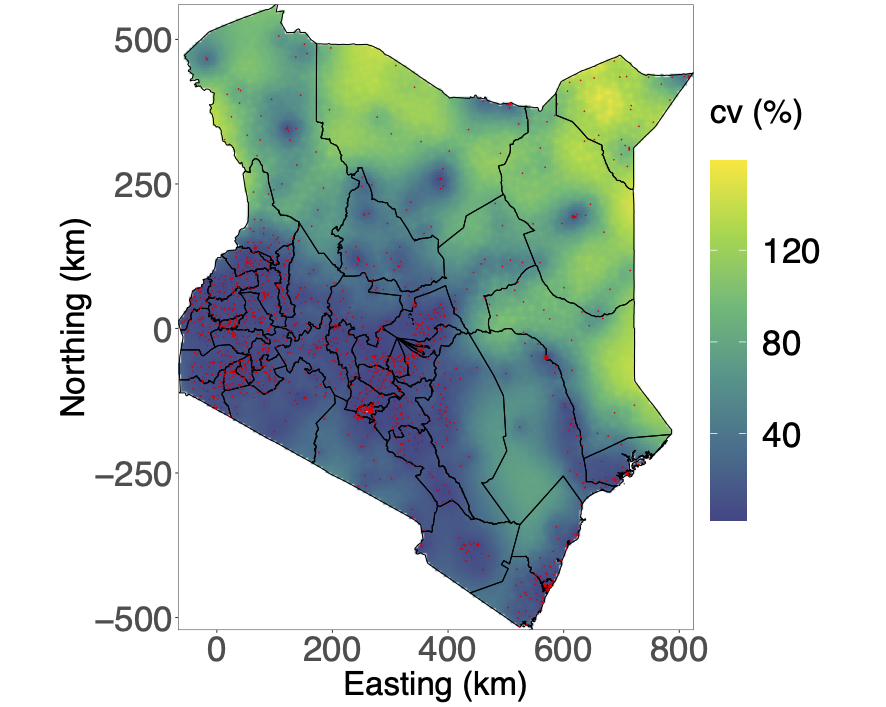}
\caption{Predicted posterior expectations (``pred.'') for the probabilities of using any contraceptive method (left) and the corresponding coefficients of variation (CV) (right) for Model-S. The red points indicate the (jittered) locations of the $C=1,583$ clusters in Kenya.\label{fig:estimatesStandard}}
\end{figure}

\begin{figure}
  \begin{subfigure}[t]{.45\textwidth}
    \centering
    \includegraphics[width=\linewidth]{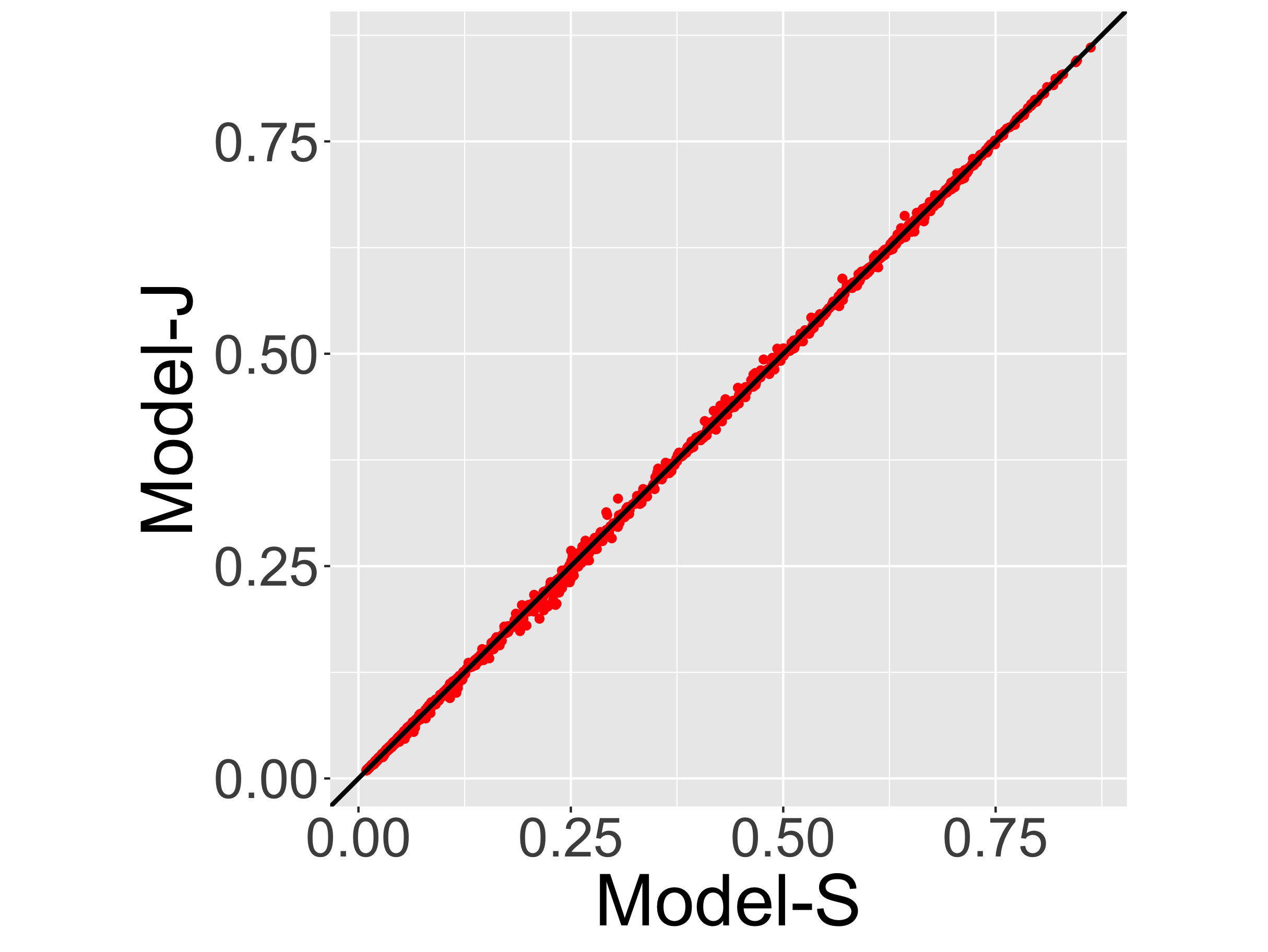}
    \caption{Scatter plot of predicted posterior expectations for the probabilities of using any contraceptive method with Model-S versus Model-J\label{fig:scatterPred}}
  \end{subfigure}
  \hfill
  \begin{subfigure}[t]{.45\textwidth}
    \centering
    \includegraphics[width=\linewidth]{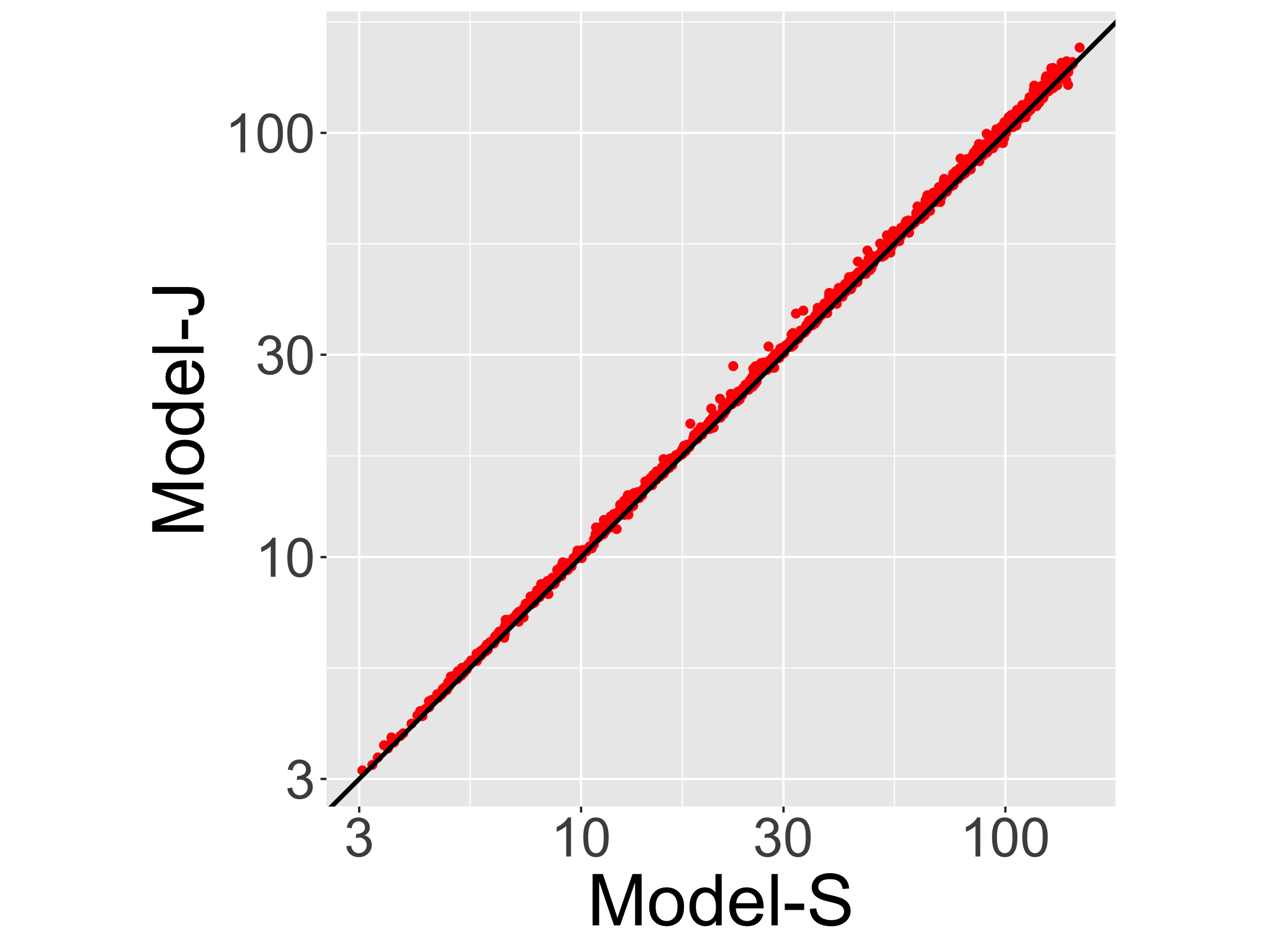}
    \caption{Scatter plot for the coefficient of variation values of the predictions (in log scale) with Model-S versus Model-J\label{fig:scatterCV}}
  \end{subfigure}
  \caption{Scatter plots of the predictions and corresponding uncertainty\label{fig:scatter}}
\end{figure}

Figure \ref{fig:estimatesStandard} shows the predicted posterior expectations for the probabilities of using any contraceptive method and the corresponding coefficients of variation (CV) for KDHS2014 contraceptive usage data with Model-S. Similar figures for Model-J are shared in Section 2 of  ``Fast geostatistical inference under positional uncertainty: Analysing DHS household survey data” paper. Figure \ref{fig:scatter} shows the comparison of the predicted posterior expectations for the probabilities of using any contraceptive method and the coefficient of variations that are obtained from Model-S and Model-J, by using KDHS2014 contraceptive usage data. Coefficient of variation values are slightly higher for Model-J compared to Model-S, while the predicted posterior expectations from both models are very similar to each other, as it is also mentioned in Section 4 of the main manuscript. 

Table \ref{fig:parEst} shows the parameter estimates and corresponding 95\% intervals for KDHS2014 contraceptive usage data with Model-J. The corresponding values using Model-S are shown in parantheses.

\begin{table}
\caption{Parameter estimates and corresponding 95\% intervals\label{fig:parEst}} 
\centering
\begin{tabular}{|c|c|c|c|c|c|}
  \hline
 & \textbf{Median} & \textbf{Lower} & \textbf{Upper} & \textbf{Length}\\
 \hline
$\beta_{0}$ & -1.78\emph{(-1.76)} & -2.61\emph{(-2.60)} & -0.97\emph{(-0.93)} & 1.64\emph{(1.66)}\\
$\rho$ & 183\emph{(188)} & 143\emph{(147)} & 233\emph{(241)} & 90\emph{(94)}\\
$\sigma^{2}_{SF}$ & 1.74\emph{(1.72)} & 1.43\emph{(1.42)} & 2.11\emph{(2.09)} & 0.68\emph{(0.67)}\\
   \hline
\end{tabular}
\end{table}

\section{Technical Derivation of Numerical Integration Procedure \label{sec:numInt}} 
If we take integration points in each ring to be angularly equidistant, and represent the area associated with the $ijk$-th integration point (for observation $i$, integration ring $j$, and the $k$-th integration point in the ring) as,
$$A_{ijk} = \{ \boldsymbol{s}_i + (r\cos a , \ r\sin a )^T: r_{i(j-1)} \leq r < r_{ij}, \  a_{ij(k - 1)} \leq  a  <  a_{ijk} \},$$
where $r_{i0}$ is taken to be 0 for all $i$, and $r_{iJ^i}$ is $L_i$, then the weights depend on the probability mass of the jittering distribution in each $A_{ijk}$. We take the integration area boundaries as equispaced, 
$$ a_{ijk} = \begin{cases}
\frac{2 \pi}{m_{ij}} (k - 1) + \frac{\pi}{m_{ij}}, & j \ \mbox{mod} \ 2 = 1, \ j \geq 5 \\
\frac{2 \pi}{m_{ij}} (k - 1), & \text{otherwise,}
\end{cases} $$
where $\frac{\pi}{m_{ij}}$ intersperses the integration points for every other ring based on $m_{ij}$, the number of integration points for observation $i$ and ring $j$. Now that each $a_{ijk}$ has been specified for $i=1,\ldots,n$, $j=1,\ldots,J^i$ given $J^i$ the number of integration rings for observation $i$, and $k=1,\ldots,m_{ij}$, the probability mass of the jittering distribution in $A_{ijk}$ and therefore the integration point weights depend only on the choice of the radii $r_{ij}$. 
Since the jittering density distribution in \eqref{eq:obsLoc} in the main manuscript is radially symmetric, interspersing the points along each ring does not influence the integration weights. Our choice of the $r_{ij}$ will depend on whether the observed cluster is urban or rural.

\begin{comment}
Before proceeding, it is important to note that, while \eqref{eq:obsLoc} in the main manuscript describes the jittering process density in $\mathbb{R}^2$, the density simplifies to a uniform distribution when reparameterized in radial coordinates on $[0, 2\pi] \times [0, L_i]$. This will make selecting the ring radii significantly simpler.
\end{comment}

For urban clusters, the jittering process density is continuous on the support of the density, unlike for the rural clusters. We choose the radii, $r_{ij}$, for any fixed urban observation $i$ so that the integration weights are equal for each of the integration points. If the prior density $\pi(\boldsymbol{s}_i^*)$ is constant over the support of $\pi(\boldsymbol{s}_i \vert \boldsymbol{s}_i^*)$, then $\pi(\boldsymbol{s}_i \vert \boldsymbol{s}_i^*)$ being uniform when represented in radial coordinates on $[0, 2\pi] \times [0, L_i]$ implies setting $r_{ij} = \frac{L_i \sum_{j'=0}^j m_{ij'}}{\sum_{j'=1}^{J^i} m_{ij'}}$ results in equal urban integration weights in \eqref{eq:numericalIntegral} in the main manuscript, with, 
$$\lambda_{ijk} \propto \frac{r_{ij} - r_{i(j-1)}}{L_i} \frac{a_{ijk} - a_{ij(k - 1)}}{2 \pi} \pi (\boldsymbol{s}_{ijk}^*),$$
so that $\lambda_{ijk} \propto \frac{r_{ij} - r_{i(j-1)}}{L_i} \frac{a_{ijk} - a_{ij(k - 1)}}{2 \pi}$ if $\pi (\boldsymbol{s}_{ijk}^*)$ is constant.

If $\boldsymbol{s}_i$ is rural, there is a discontinuity in $\pi (\boldsymbol{s}_i \vert \boldsymbol{s}_i^*)$ where $||\boldsymbol{s}_i - \boldsymbol{s}_i^*|| = L_i'$ for discontinuity radius $L_i'$ due to the fact that there is a $0.01$ probability of rural points having a larger maximum jittering distance. We therefore define `inner' and `outer' rings with $J^i = J^i_{\text{inner}} + J^i_{\text{outer}}$, where the inner rings and outer rings are inside and outside of the discontinuity radius respectively. For rural DHS spatial locations, $r_{iJ^i_{\text{inner}}} = L_i' = 5$ and $r_{iJ^i} = L_i = 10$. We choose the inner and outer ring radii so that the integration points in the inner and outer rings have equal weights respectively, so that:
$$ r_{ij} = \begin{cases}
\frac{\sum_{j'=1}^{j} m_{ij'}}{\sum_{j'=1}^{J^i_{\text{inner}}} m_{ij'}} L_i', & 1 \leq j \leq J^i_{\text{inner}} \\
L_i' +  \frac{\sum_{j'=J^i_{\text{inner}} + 1}^{j} m_{ij'}}{\sum_{j'=J^i_{\text{inner}} + 1}^{J^i} m_{ij'}} (L_i - L_i'), & J^i_{\text{inner}} < j \leq J^i.
\end{cases}$$
These ring radii result in the following rural integration weights: 
\begin{align}
\lambda_{ijk} &\propto 
\begin{cases}
\frac{r_{ij} - r_{i(j-1)}}{m_{ij}} \left( \frac{99}{100} \frac{1}{r_{J^i_{\text{inner}}}} + \frac{1}{100} \frac{1}{L_i} \right), & 1 \leq j \leq J^i_{\text{inner}} \\
\frac{r_{ij} - r_{i(j-1)}}{m_{ij}}  \frac{1}{100} \frac{1}{L_i}, & J^i_{\text{inner}} < j \leq J^i.
\end{cases} \label{eq:integrationWeightRural}
\end{align}
The $\frac{99}{100}$ and $\frac{1}{100}$ factors in the above expressions are due to rural clusters having a probability of $\frac{1}{100}$ of being displaced by up to 10 km. We set $J^i=5$ for urban points, and $J^i_{\text{inner}} = 5$ and $J^i_{\text{outer}} = 5$ for rural points. Although the rural outer ring weights are much smaller than the inner weights to the point where leaving them out and renormalizing the weights would likely not influence the predictions, and would improve computation times, we choose to include them for greater precision.

We set each integration point $\boldsymbol{s}_{ijk}^*$ to be the center of mass of $\pi (\boldsymbol{s}_i \vert \boldsymbol{s}_i^*)$ within the associated $ijk$-th integration area $A_{ijk}$, with $\boldsymbol{s}_{ijk}^* = (r_{ij}^* \cos((a_{ijk} + a_{ijk})/2), \ r_{ij}^* \sin((a_{ijk} + a_{ijk})/2))^T$, and where,
\begin{equation}
r_{ij}^* = \frac{r_{i(j-1)} + r_{ij}}{2} \frac{\sqrt{2(1 - \cos(a_{ij2} - a_{ij1}))}}{a_{ij2} - a_{ij1}},
\label{eq:displacement}
\end{equation}
for $j > 1$ (if $j=1$, then $r_{ij}^* = 0$). A derivation of \eqref{eq:displacement} is given in more detail below.

If observation $i$ is urban, we can calculate the expectation of the horizontal coordinate, say $x_{ij1}$, for the center of mass of the first integration area in ring $j$, and assuming $a_{ij0} = 0$ and $a_{ij1} = 2\pi / m_{ij}$, as follows:
\begin{align}
E[x_{ij1}] &= \int_{r_{i(j - 1)}}^{r_{ij}}  \int_{0}^{2\pi / m_{ij}} r x \frac{C_{ij}}{2 \pi D_i r} \ \mathrm{d}a \ \mathrm{d}r \nonumber \\
&= \int_{r_{i(j - 1)}}^{r_{ij}}  \int_{0}^{2\pi / m_{ij}} r \cos(a) \frac{C_{ij}}{2 \pi D_i} \ \mathrm{d}a \ \mathrm{d}r \nonumber \\
&= \int_{r_{i(j - 1)}}^{r_{ij}} r \frac{C_{ij}}{2 \pi D_i} (\sin(2\pi / m_{ij}) - \sin(0)) \ \mathrm{d}r \nonumber \\
&= \frac{C_{ij}}{2 \pi D_i} \sin(2\pi / m_{ij}) (r_{ij}^2 - r_{i(j-1)}^2) \nonumber \\
&= \frac{\sin(2\pi / m_{ij})}{2 \pi / m_{ij}} \frac{r_{ij} + r_{i(j-1)}}{2}, \label{eq:integrationPointx}
\end{align}
where $C_{ij} = \frac{2 \pi D_i}{(r_{ij} - r_{i(j-1)}) (a_{ijk} - a_{ij(k-1)})}$. Similar reasoning yields the following expectation for $E[y_{ij1}]$, where $y_{ij1}$ is the vertical coordinate of the center of mass of the first integration area for observation $i$ in ring $j$: 
\begin{equation}
E[y_{ij1}] = \frac{1 - \cos(2\pi / m_{ij})}{2 \pi / m_{ij}} \frac{r_{ij} + r_{i(j-1)}}{2}.
\label{eq:integrationPointy}
\end{equation}

We can combine the above two expectations to get the radial displacement of the center of mass of integration area $A_{ijk}$, relative to $\boldsymbol{s}_i$: 
\begin{align}
r_{ij}^* &= \sqrt{E[x_{ij1}]^2 + E[y_{ij1}]^2} \nonumber \\
&= \sqrt{\frac{\sin(2\pi / m_{ij})^2}{4 \pi^2 / m_{ij}^2} \frac{(r_{ij} + r_{i(j-1)})^2}{4} + \frac{(1 - \cos(2\pi / m_{ij}))^2}{4 \pi^2 / m_{ij}^2} \frac{(r_{ij} + r_{i(j-1)})^2}{4}} \nonumber \\
&= \frac{r_{ij} + r_{i(j-1)}}{2} \frac{m_{ij}}{2 \pi} \sqrt{\sin(2\pi / m_{ij})^2 + (1 - \cos(2\pi / m_{ij}))^2} \nonumber \\ 
&= \frac{r_{ij} + r_{i(j-1)}}{2} \frac{m_{ij}}{2 \pi} \sqrt{2 (1 - \cos(2\pi / m_{ij}))}. \label{eq:integrationPointRadius}
\end{align}
\noindent
Due to the radial symmetry of the jittering distribution under a flat prior $\pi(\boldsymbol{s}_i^*)$, we obtain $\sqrt{E[x_{ijk}]^2 + E[y_{ijk}]^2} = \sqrt{E[x_{ij1}]^2 + E[y_{ij1}]^2}$ for all $1 \leq k \leq m_{ij}$.

If observation $i$ is rural, we must use the rural jittering density taking the form,
$$\pi(r) = \begin{cases}
\frac{99}{100} \frac{C_{ij}}{2 \pi D_i r} + \frac{1}{100} \frac{C_{ij}'}{2 \pi D_i' r}, & 0 < r \leq L_i' \\
\frac{1}{100} \frac{C_{ij}'}{2 \pi D_i' r}, & L_i' < r \leq L_i\\
0, & \text{otherwise,}
\end{cases}$$
for $C_{ij} = \frac{2 \pi D_i}{(r_{ij} - r_{i(j-1)}) (a_{ijk} - a_{ij(k-1)})}$ and $C_{ij}' = \frac{2 \pi D_i'}{(r_{ij} - r_{i(j-1)}) (a_{ijk} - a_{ij(k-1)})}$. We can then calculate the expected horizontal coordinate of the integration area with respect to the rural jittering density in the same way as for the urban density: 

$$E[x_{ij1}] = \int_{r_{i(j - 1)}}^{r_{ij}}  \int_{0}^{2\pi / m_{ij}} r x  \left ( \frac{99}{100} \frac{C_{ij}'}{2 \pi D_i' r} + \frac{1}{100} \frac{C_{ij}}{2 \pi D_i r} \right ) \ \mathrm{d}a \ \mathrm{d}r. $$

Since \eqref{eq:integrationPointRadius} does not depend on $D_i$, we reach the same result for rural as for urban integration points for `inner' integration area $A_{ijk}$:

\begin{align*}
E[x_{ij1}] &= \frac{99}{100} \int_{r_{i(j - 1)}}^{r_{ij}}  \int_{0}^{2\pi / m_{ij}} r x \frac{C_{ij}'}{2 \pi D_i' r} \ \mathrm{d}a \ \mathrm{d}r + \frac{1}{100} \int_{r_{i(j - 1)}}^{r_{ij}}  \int_{0}^{2\pi / m_{ij}} r x \frac{C_{ij}}{2 \pi D_i r} \ \mathrm{d}a \ \mathrm{d}r \\
&= \frac{99}{100} \frac{r_{ij} + r_{i(j-1)}}{2} \frac{m_{ij}}{2 \pi} \sqrt{2 (1 - \cos(2\pi / m_{ij}))} + \frac{1}{100} \frac{r_{ij} + r_{i(j-1)}}{2} \frac{m_{ij}}{2 \pi} \sqrt{2 (1 - \cos(2\pi / m_{ij}))} \\
&= \frac{r_{ij} + r_{i(j-1)}}{2} \frac{m_{ij}}{2 \pi} \sqrt{2 (1 - \cos(2\pi / m_{ij}))}.
\end{align*}
Similar lines of reasoning show that the above expression for $E[x_{ij1}]$ holds even for `outer' integration areas, and that \eqref{eq:integrationPointy} and \eqref{eq:integrationPointRadius} also hold for rural integration areas (both inner and outer).

Table \ref{tab:integrationInfo} gives the radial displacement, number of integration points, and integration weights (uncorrected for potential administrative boundary effects) as a function of $j$, the ring index.

\begin{table}
\caption{For each numerical integration ring, the displacement, number, and weights of the individual integration points. Weights here have not been corrected for edge effects, and have been normalized to sum to 1. Displacements are scaled to match the DHS jittering distribution.}
\centering
\resizebox{\linewidth}{!}{\begin{tabular}{lcccc}
& \textbf{Ring Number} & \textbf{Displacement (km)} & \textbf{Number of Points} & \textbf{Integration Weights} \\ 
  \hline
\multirow{5}{*}{\textbf{Urban}} & 1 & 0.00 & 1 & 0.0164 \\ 
& 2 & 0.28 & 15 & 0.0164 \\ 
& 3 & 0.76 & 15 & 0.0164 \\ 
& 4 & 1.25 & 15 & 0.0164 \\ 
& 5 & 1.74 & 15 & 0.0164 \\ 
\hline
\multirow{10}{*}{\textbf{Rural}}& 1 & 0.00 & 1 & 0.0163 \\ 
& 2 & 0.69 & 15 & 0.0163 \\ 
& 3 & 1.91 & 15 & 0.0163 \\ 
& 4 & 3.13 & 15 & 0.0163 \\ 
& 5 & 4.35 & 15 & 0.0163 \\ 
& 6 & 5.46 & 15 & 0.0001 \\ 
& 7 & 6.45 & 15 & 0.0001 \\ 
& 8 & 7.45 & 15 & 0.0001 \\ 
& 9 & 8.44 & 15 & 0.0001 \\ 
& 10 & 9.43 & 15 & 0.0001\\
   \hline
\end{tabular}}

\label{tab:integrationInfo}
\end{table}

\end{appendices}

\end{document}